\documentclass[preprint,12pt,authoryear]{elsarticle}

\usepackage{amssymb}
\usepackage[a4paper, outer=2.7cm, top=3cm, bottom=3cm]{geometry}
\usepackage{graphicx}

\usepackage{array}
\usepackage[utf8]{inputenc}
\usepackage{fancyhdr}
\usepackage[printonlyused]{acronym}
\usepackage{comment}
\usepackage{listings}
\usepackage{amsmath}
\usepackage{float}
\usepackage{caption}
\usepackage{subcaption}
\usepackage{eurosym}

\DeclareUnicodeCharacter{20AC}{\euro}

\journal{---}

\begin{document}

\begin{frontmatter}



\title{Analysis of Parallel Boarding Methods in a Multi-Aisle Flying Wing Aircraft}


\author[inst1,inst2]{Emil Ryd}

\affiliation[inst1]{organization={UWC Red Cross Nordic},
            country={Norway}}
\affiliation[inst2]{organization=
            {New College, University of Oxford},
            country={United Kingdom}}
\affiliation[inst3]{organization=
            {University of California, Berkeley},
            country={USA}}
\affiliation[inst4]{organization=
            {Hong Kong University of Science and Technology},
            country={Hong Kong}}
\author[inst1,inst3]{Vihaan Khandelwal}
\author[inst1,inst4]{Hayden So}
\author[inst6]{Jason H. Steffen}

\affiliation[inst6]{organization={Department of Physics and Astronomy, University of Nevada, Las Vegas},
            country={USA}}

\begin{abstract}
We examine the speed of different boarding methods in a proposed Flying Wing aircraft design with four aisles using an agent-based model. We study the effect of various passenger movement variables on the boarding process. We evaluate the impact of these factors on the boarding time when the boarding process runs sequentially and in parallel with the aisles of the Flying Wing layout. Then, we analyze the impact of an increase in the number of aisles on the relative speed of all boarding methods and conclude that methods utilizing boarding of the separate aisles simultaneously (parallel boarding) converge to the fastest boarding time given by the Steffen method.  With parallel boarding of the aisles the relative advantage of the Steffen method compared to Windows-Middle-Aisle (WMA) or Back-to-front boarding decreases, from being 1.6-2.1 times as fast to being approximately equal for our fiducial Flying Wing seating arrangement. Standard methods such as Back-to-front or WMA are about twice as fast to board a four-aisle Flying Wing plane, compared to a single-aisle aircraft with the same number of passengers. We also investigate the difference between the optimal approach to parallel boarding, where consecutive passengers always enter separate aisles, and a less optimal but more practical approach. The practical approach is only up to 1.06 times slower than the optimal, meaning that the advantages of parallel boarding can be utilized without resorting to impractical boarding methods. Hence, the introduction of multiple aisles into aircraft seating design offers the prospect of significantly decreasing the boarding time for passengers, without the introduction of inconvenient boarding methods.

\end{abstract}



\end{frontmatter}



\newpage
\section{Introduction}
\label{sec:sample1}


Starting with \citet{Marelli1998}, and subsequently \citet{Belgium2002}, the boarding process has been a prominent direction of optimization research as it plays a fundamental role in the collective experience of the crew and passengers. See \citet{AB} for a comprehensive review. Its close connection to the passenger experience as well as the fact that airplane boarding has been shown to be on the critical path of turnaround time \citep{NEUMANN2019} merits its emphasis in the literature. And while the estimated cost of extended boarding times vary significantly depending on the details of the flight, \citet{Cook2011EuropeanAD} state a high-level estimate of \euro 100 per minute. Adding up this cost over the course of a year across all flights, reducing boarding times by just incremental amounts could still lead to drastic increases in profits or revenue for an airline \citep{AB}. \cite{AB} emphasize the dual objectives of boarding time and passenger comfort as the two main metrics for evaluating boarding procedures, the former quantitatively and the latter qualitatively, and highlight the need for boarding methods to take both of these objectives into account \citep{AB}. \\

Following the initial study by \citet{Marelli1998}, research into boarding has primarily been done along three separate strands: computer simulations, physical experiments, and theoretical models \citep{Belgium2002, STEFFENExperiment, GeneralStudy, AB}. Studies have often complemented each other, e.g. theoretical models being validated by results from computer simulations, and using parameters derived from experiment \citep{Bachmat2009, Erland2022}. \\  

Computer simulations were the main tool in the first attempts at approaching the problem of optimizing boarding procedures, and they have been widely employed ever since \citep{Willamowski2022}. These studies typically investigate a few well-defined boarding strategies and compare their relative speeds given model parameters \citep{AB}. These boarding strategies (excluding random boarding) may in general be sub-divided into two groups, \emph{by-group} and \emph{by-seat} \citep{AB}. By-group strategies split passengers into a small number of different groups which order in a specified sequence, but where the internal order of the group is random, e.g. Back-to-front boarding. Meanwhile, in by-seat strategies every single passenger is ordered in a specified sequence in which to board the aircraft \citep{Willamowski2022}. \\

Early computer simulation studies into boarding found that popular by-group strategies such as Back-to-front in fact performed worse than random boarding, and that by-seat strategies such as Window-Middle-Aisle (WMA, or WilMA, also known as Outside-In) often executed significantly faster \citep{Marelli1998} \citep{Belgium2002}. Using a cell-based model, \citet{Ferrari2005} examined the effect of disturbances on boarding procedures, and found that Back-to-front improved with more late-arriving passengers. \citet{vandenBriel2005} introduced the \emph{reverse-pyramid} boarding method which combined both Back-to-front and WMA to yield even slightly faster boarding \citep{vandenBriel2005}. \\

Seeking to find an optimal boarding method, \citet{Steffen2008} theorized that the key aspect was maximizing the number of passengers stowing their luggage simultaneously, and thus the key to a fast boarding method was minimizing aisle interferences. He introduced a novel optimization method using a Markov Chain Monte Carlo algorithm, and thus managed to find a boarding method which was optimal given negligible walking time between aircraft entrance and passenger seats. The results were later corroborated in an experimental study \citep{STEFFENExperiment}, and to this day this remains the fastest boarding method given these parameters \citep{KIERZKOWSKI2017}. Nevertheless, the by-seat nature of this boarding method means that it is infeasible to implement in practice and would possibly lead to a worse passenger experience. For this reason, \citet{Steffen2008} also proposed a modified approach which would be more practical yet still provide efficient boarding. \\

Starting with \citep{vandenBriel2005}, researchers have also created analytical models to study the boarding question. Often using concepts from theoretical and statistical physics and making some simplifying assumptions such as infinite passenger walking speed, these models typically give probabilistic bounds for boarding times, i.e. when the number of passengers tend to infinity \citep{Willamowski2022}. Despite only explicitly giving information about the asymptotic case, the analytical results from these models have been insightful and also shown to corroborate with computer simulation \citep{AB}. Notably, \citet{Bachmat2009} uncovered parallels between analytical treatment of the boarding problem and Lorentzian space-time geometry, and found agreements between their closed-form expressions and the computer simulations of \citet{Belgium2002}.\\

More recently, both analytical and computational studies have explored \textit{dynamic} boarding procedures, where the passengers are split based on individual properties not related to their seating, such as carry-on luggage or walking speed \citep{Bachmat2019, NewMethod, MILNE2016, NOTOMISTA2016, Audenaert2009}. Specifically, \citet{TANG2012} find that by taking dynamic variables such as luggage and walking speed into account, interferences and congestion can be eliminated from the boarding procedure. Other studies have found that letting the faster passengers board last could yield faster boarding times \citep{Erland2019, Erland2021}. While these results may be of theoretical interest to optimization literature, their significance for airline management is not evident yet as it is still unclear if the individual passenger boarding speed can be reliably determined \textit{a priori}, although the number of luggage items may be a good enough predictor, as examined in \citep{Nugroho2022}. \\

As the field of boarding research has progressed, researchers have investigated more parameters in their models, and there have also been larger empirical studies connecting model parameters to real-world data on passengers \citep{Schultz2018_58, Schultz2018a}. Additionally, some recent studies explore the impact of aircraft design features on boarding time \citep{Schultz2017, Fuchte2014}. \\

\citet{Bazargan2011} was the first to consider boarding in a wide-body aircraft, and used a linear integer programming model to find an optimal boarding method for a Boeing 767. Other works on wide-body aircraft include \citep{Schultz2013, Giitsidis2016, IEEEBoarding}. These papers adapt existing boarding models to two-aisle configurations, yet they do not explicitly examine the effect of parallelizing the boarding procedure which occurs in multi-aisle aircraft. Effectively, these aircraft can have two boarding procedures running simultaneously, yielding substantial improvements in speed. \citet{Bazargan2016} explore a similar phenomenon in their study of boarding an airplane with multiple entrances.\\

With the introduction of multi-aisle aircrafts of four or more aisles like the 'Flying-V' \citep{Benad2015}, which researchers at TU Delft and elsewhere have been working on since 2013, the consideration of parallel boarding has for the first time gained practical importance. Flying Wing aircraft like the one proposed by the group at TU Delft are still in the prototype stage, they offer a promising alternative for long-haul flights, and KLM and Delft have already successfully tested a scale prototype version \citep{KLM} - see section \ref{sec:flywing} for a more detailed description of the Flying Wing aircraft. Yet, there is no definitive study on boarding in such aircraft or the effect on boarding of increasing the number of aisles of an aircraft. In this paper, we analyze the effects of multiple aisles on boarding as a question on its own, and then focus on a generic Flying Wing design to quantitatively evaluate the benefits of a multi-aisle aircraft.\\

One concern raised in the course of this research is that of the critical path. Typically, boarding procedure is not considered to be as critical for long haul flights \citep{NEUMANN2019}, for which the Flying Wing is mainly considered. Nevertheless, we believe this investigation still to be relevant as a) airlines are interested in decreasing boarding time beyond the critical turnover time to accommodate for unexpected delays in boarding \citep{AB} b) a shorter boarding procedure could improve passenger comfort and experience  c) considerations of parallel boarding in general may give insights into boarding procedures at large, specifically for two-aisle aircraft and single-aisle aircraft with multiple entrances.\\

Using an agent-based model developed in Repast Simphony Java \citep{repastsimphony}, we investigate the effect of changes to each variable and ultimately provide insight into the most salient effects for the boarding process for wide-body aircraft.  The relationship between structural design (number of aisles, number of rows in each aisle and number of seats per row) and boarding time has not been thoroughly investigated for the Flying Wing format.  We fill this gap by exploring different layouts for the Flying Wing designs and compare them to narrow-body and two-aisle aircraft with the same number of seats.\\

This paper is outlined as follows: Section \ref{sec:flywing} details the background of the Flying Wing aircraft and focuses on the structure and boarding methods that will be explored. Section \ref{sec:methodology} concerns the methodology and outlines the structure and details of our model. Section \ref{sec:human-variable-analysis} describes the results from the sensitivity analysis of human variables - referring to variables relating to the passengers properties, e.g. stowage time or walking speed, and Section \ref{sec:analysis-of-multiple-aisles} is dedicated to analysis of the effect on boarding of adding more aisles into an aircraft. We present our results in Section \ref{sec:results}, and in section \ref{sec:disc-management} we discuss the implications for airline management. Section \ref{sec:conclusion} contains our concluding points.

\section{Aircraft Design: Flying Wing}\label{sec:flywing}

\subsection{Description of seating plan}
In this investigation of the Flying Wing aircraft, we take as a model the design provided by the International Mathematical Modeling Challenge 2022, as visualized in Figure 1. We adopt a simple 4 aisle seating plan consisting of 336 seats. This layout gives a passenger count that is split among the different seating blocks as follows: 42-84-84-84-42. This layout serves as our fiducial model for this investigation.\\

\begin{figure}[H]
    \centering
    \includegraphics[scale=0.5, angle=270]{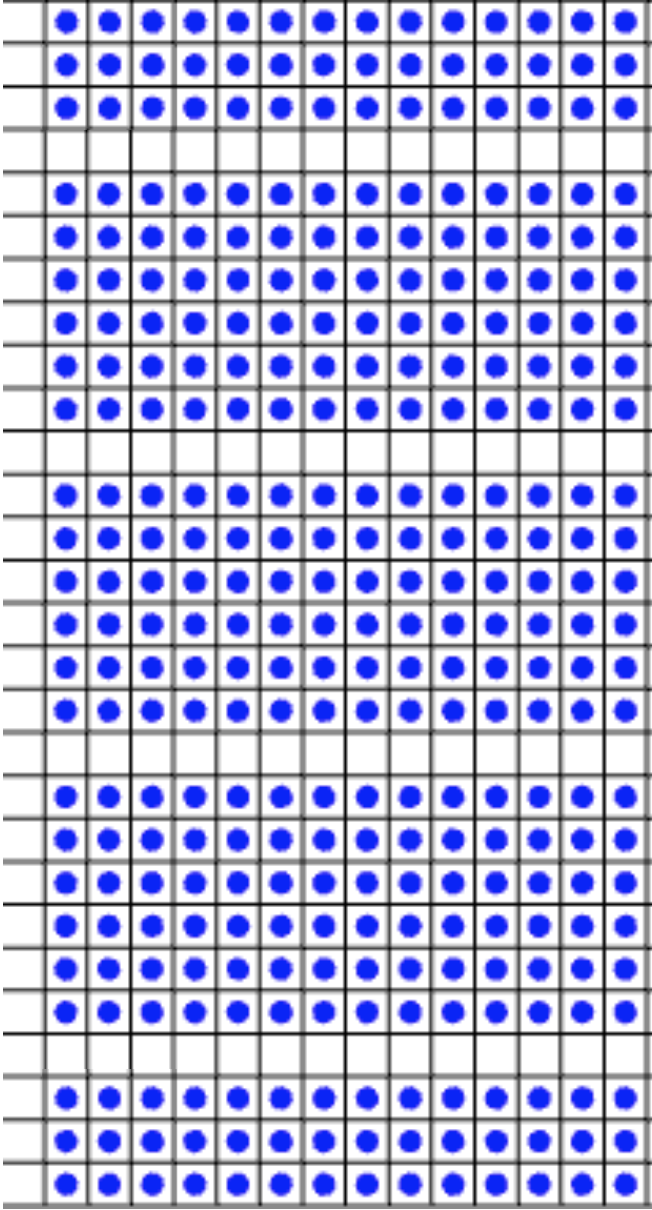}
    \caption{Flying Wing Basic Seat Plan}
    \label{fig:flying-wing-basic-structure}
\end{figure}

In 2021, researchers at TU Delft investigated a new floor plan concept of a 'Flying-V' aircraft with seats angled at 26 degrees to the direction of flight \citep{Wamelink2021}. This model consists of 4 aisles: with 2 aisles on each side of the 'V', a design very similar to the one explored in this paper.  In our fiducial setup, we have sought to replicate their suggested seating plan. Related to the seating plan, the research group at TU Delft has also found a Flying Wing aircraft could yield benefits relating to ergonomics and the comfort of the passenger \citep{Wamelink2021}.

\subsection{Boarding methods}\label{subsec:boarding-methods}
\begin{figure}[H]
    \centering
    \includegraphics[scale=0.28]{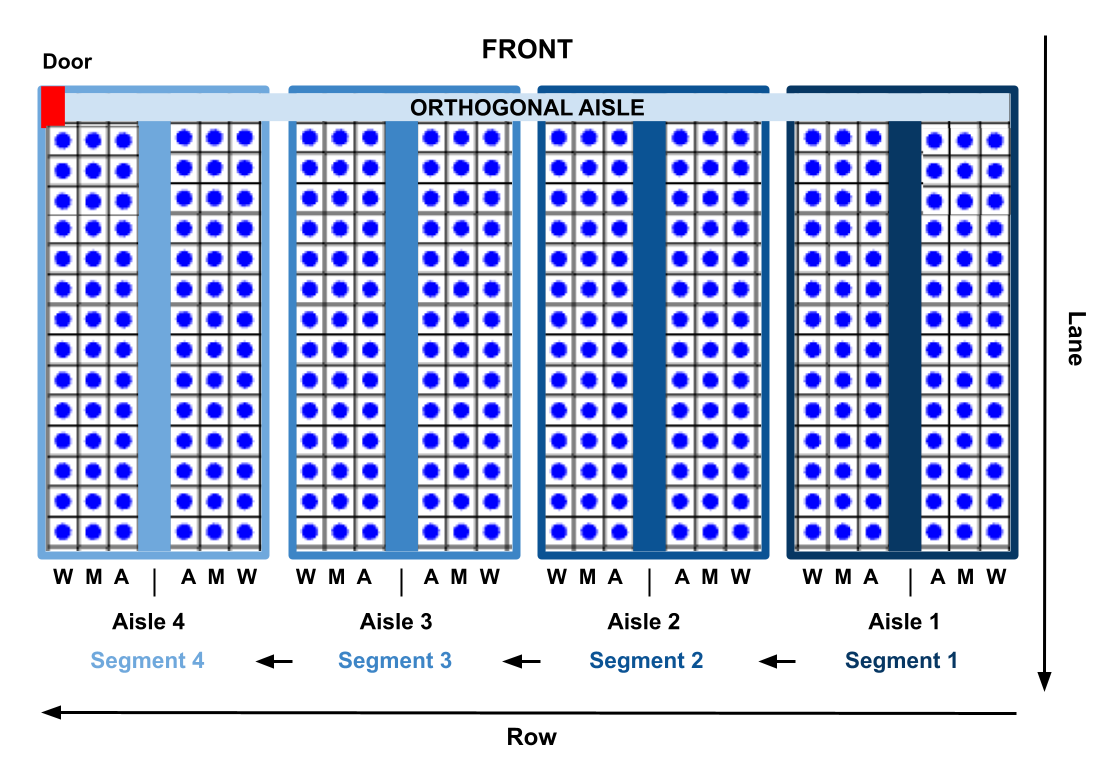}
    \caption{Breakdown of Flying Wing Structure}
    \label{fig:flying-wing-basic-structure}
\end{figure}

Figure \ref{fig:flying-wing-basic-structure} shows the layout for our nominal Flying Wing model.  Passengers enter through the door on the left and travel through the orthogonal aisle to reach their designated aisle and find their seat. For the sake of simplicity, the Flying Wing with four aisles can be treated as four separate narrow-body aircraft such as a Boeing 737 or an Airbus A320 (denoted as 'segments' in this paper), except that the collection of seats on the very left and right have nine seats less.\\

We test several boarding methods using this layout for the Flying Wing interior. These methods include:

\begin{enumerate}
    \item \textbf{Window, Middle, Aisle (WILMA/WMA):} Each lane is completed one by one starting with the one of the window lanes, completing the other window lane, and continuing this pattern with the middle and aisle lanes 
    \item \textbf{Back-to-front:} The plane is divided into groups by rows, and the group that is furthest back boards first.  After all of the passengers in that boarding group have entered, the group that is second-furthest back starts, and so on. We typically divide the airplane into three groups.
    \item \textbf{Random:} Completely random boarding dictated by the agent-based model (explained further in Section \ref{sec:methodology}).
    \item \textbf{Steffen}: As suggested by Steffen in his paper, in the one-aisle aircraft, the first person who boards has their seat in the furthest back window seat on the right. The second person's seat is the window seat two rows in front of the first person. This pattern of sitting in right window seats in the odd rows continues, and then the same pattern is done on the left side. Next, the right window seats in even rows board. This pattern continues until all the lanes are filled. 
\end{enumerate}

We choose to analyze Random, WMA, and Back-to-front as these are the methods most commonly employed by airlines \citep{Milne2020Covid}, and are thus of most practical importance. More complicated methods such as the reverse pyramid \citep{vandenBriel2005} are not typically employed by airlines due to their impractical complexity leading to difficulties in implementation and loss of passenger experience. Finally, we choose to also evaluate the Steffen method, which albeit impractical was shown to be optimal under certain conditions, hence providing a fastest possible boarding time to be used when comparing boarding methods. To focus on the structural aspects of who to utilize parallel aisles in boarding, we also omit any dynamic boarding methods, especially as these also suffer from difficulty in implementation and hence are not currently used by airlines.\\

We generalize the boarding methods listed above to work for multi-aisle aircraft. As part of our naming convention, we use the words \textit{Parallel} and \textit{Sequential} to detail the boarding methods. \textit{Parallel} is when successive passengers filter through successive aisles in the aircraft---boarding each segment of the airplane simultaneously.  \textit{Sequential} is a pattern where passengers board all of segment 1 first, followed by all of segment 2, segment 3, and segment 4.  Below is a list of the boarding methods we generalize to and test in the Flying Wing model.

\begin{enumerate}
    \item\textbf{Random:} As mentioned above, completely random boarding dictated by the agent-based model (explained further in Section \ref{sec:methodology}).
    \item \textbf{Parallel Random:} Random boarding with the condition that the first passenger has a random seat in segment 1, the next having a random seat in segment 2 and so on.
    
    \item \textbf{Parallel Back-to-front in three groups (Parallel BTF):} Looking at figure \ref{BTF-P}, each segment is split into three groups. In this method, the first passenger boards a random seat in the back group of segment 1. This is followed by passenger boarding a random seat in the back group of segment 2 and so on. When the back group seats are filled, it then completes the middle group of segment 1, 2, 3 and 4 in that order.
  
    \item \textbf{Parallel Front-to-back in three groups (Parallel FTB):} This is the same as Parallel BTF, except the front group is completed first, so essentially in reverse order.

    \item \textbf{Parallel Steffen:} Like Steffen's method on a narrow-body aircraft, this method boards passengers via Steffen's method but by having the next passenger go to the next segment. Passenger 1 would go to Segment 1 and go to their seat according to Steffen's method, passenger 2 will go to segment 2 and do the same. In other words, passengers 1, 5, 9, 13... will be seated in segment 1.

    \item \textbf{Parallel Window Middle Aisle (WMA) Random:} This method feeds in the first passenger to segment 1, where they could be in any of the window seats. Second passenger enters into segment 2 and could find themselves in any of the window seats. When all the window seats are occupied, the middle lane seats are filled up at random, with first passenger to segment 1 and second to segment 2, and so until all seats are filled.

    \item \textbf{Parallel Worst Method:} This method details the worst parallel method. The first passenger takes the right aisle seat at the front row of segment 1. The second passenger takes the right aisle seat at the front row of segment 2. When this is done for all four segments, the left aisle seat at the front row is taken, starting with segment 1. After this, the right middle seat at the front row of segment 1 is taken. This process repeats. As described, this is the worst possible parallel boarding method.
  
    \item \textbf{Sequential Steffen:} This method boards Segment 1, 2, 3 and 4 via Steffen's method in that order. Essentially, this method can be treated as repeating Steffen's method four times on four separate `mini narrow-body aircraft (segments)'. 
  
    \item \textbf{Parallel Sequential Steffen:} This method combines Steffen with parallel and sequential properties; Instead of feeding in passengers one by one into each segment and making them complete boarding by Steffen's method, the odd rows on the right `W (window)' lane of segment 1 is boarded first, followed by odd rows of the right `W' lane of Segment 2. Once all the odd rows on the right are boarded, the odd rows of the left `W' lane is boarded and once this is done, the even rows of the `W' lane is boarded.
 
    \item \textbf{Parallel Sequential Window Middle Aisle WMA Random:} This method boards the right `W' lane of segment 1 first, followed by the right `W' lane of segment 2 and so on. Once the right `W' lanes are done it moves onto the left `W' lanes, starting from segment 1. The random aspect comes in from the fact that the passengers could be anywhere within the lane.

\begin{figure}[H]
\centering
    \subfloat[Parallel BTF 3 Groups \label{BTF-P}]{\includegraphics[width=0.47\linewidth]{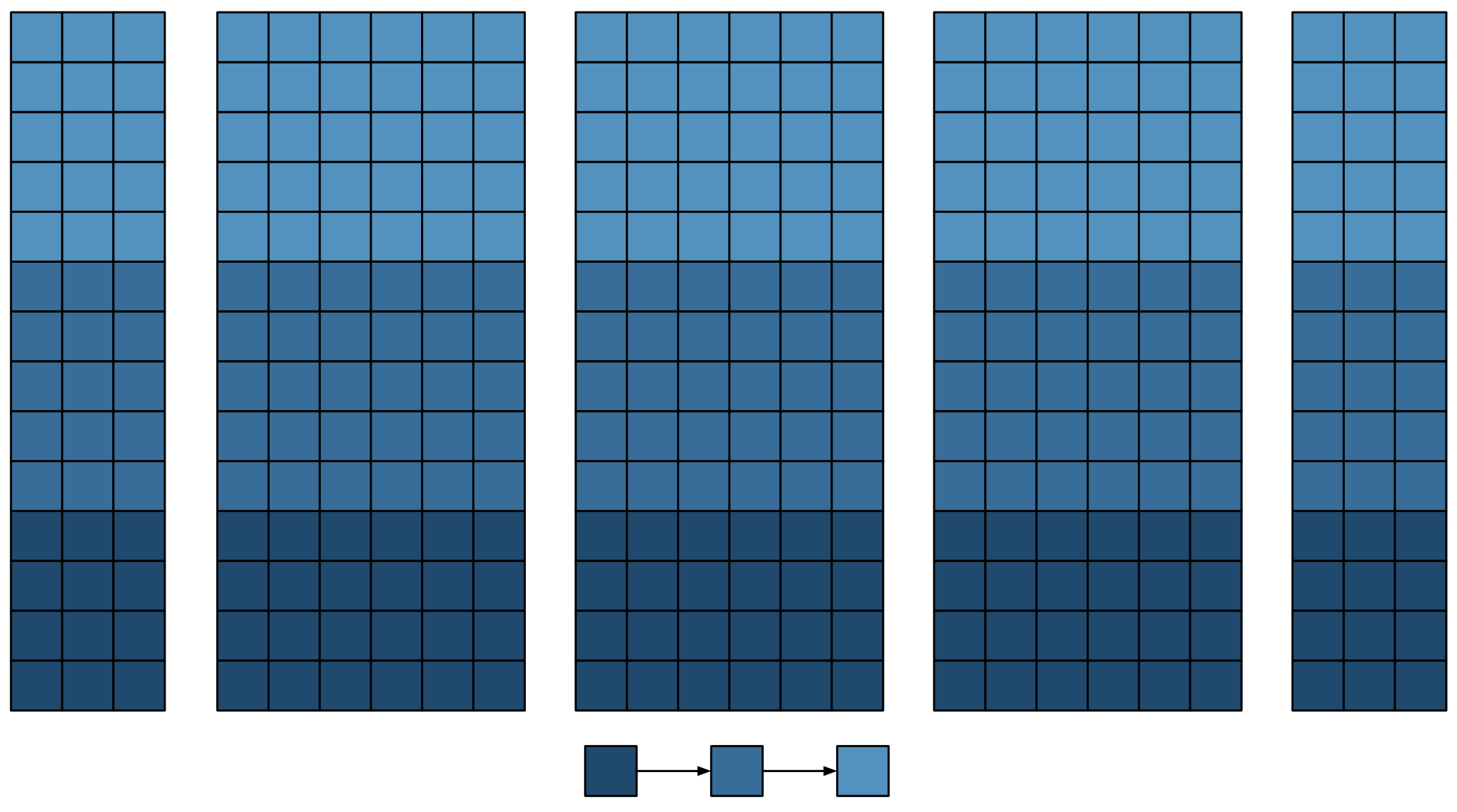}}
\hfil
    \subfloat[Parallel WMA Random \label{WMA-R}]{\includegraphics[width=0.47\linewidth]{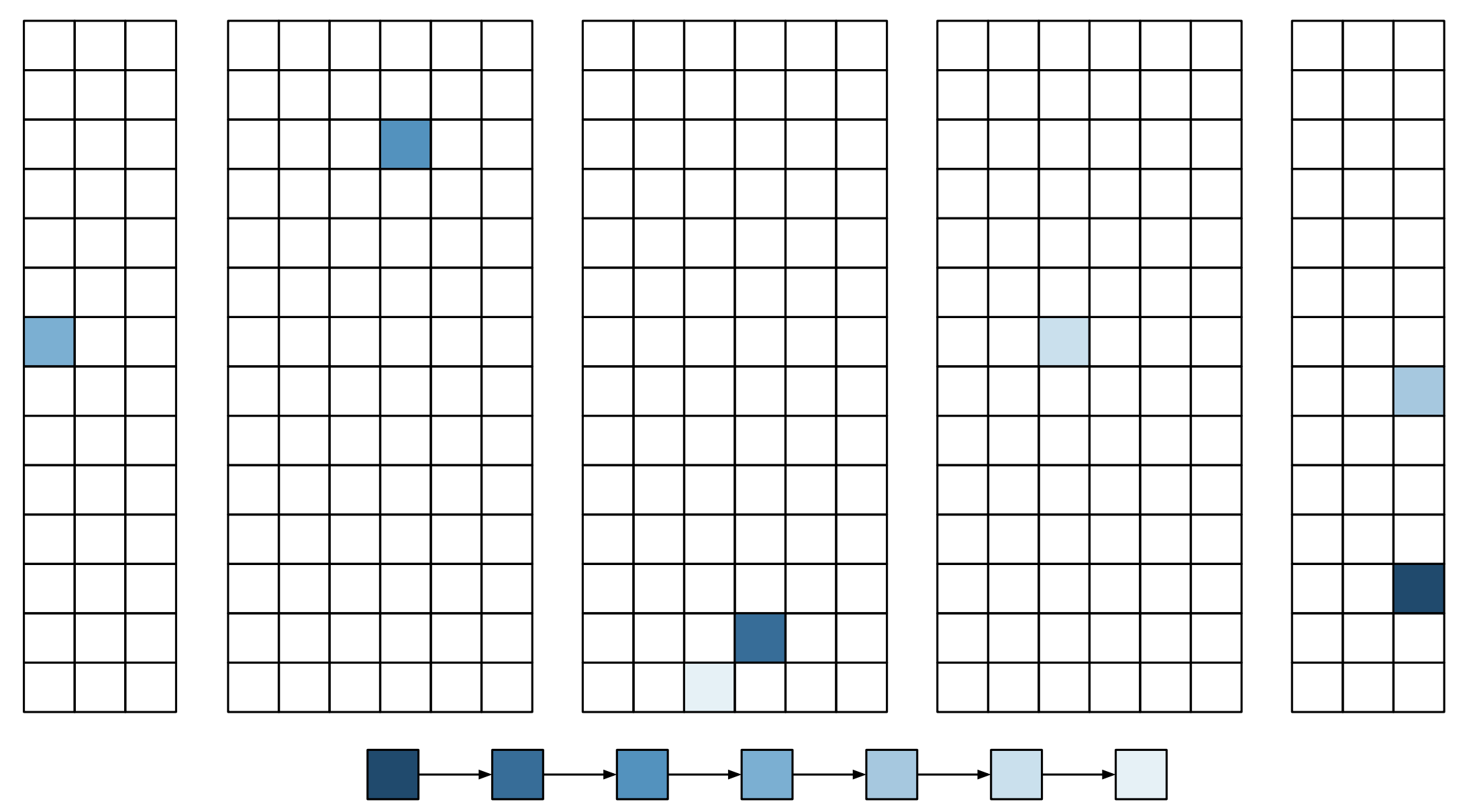}}

    \subfloat[Parallel Sequential WMA Random \label{PS-WMA}]{\includegraphics[width=0.47\linewidth]{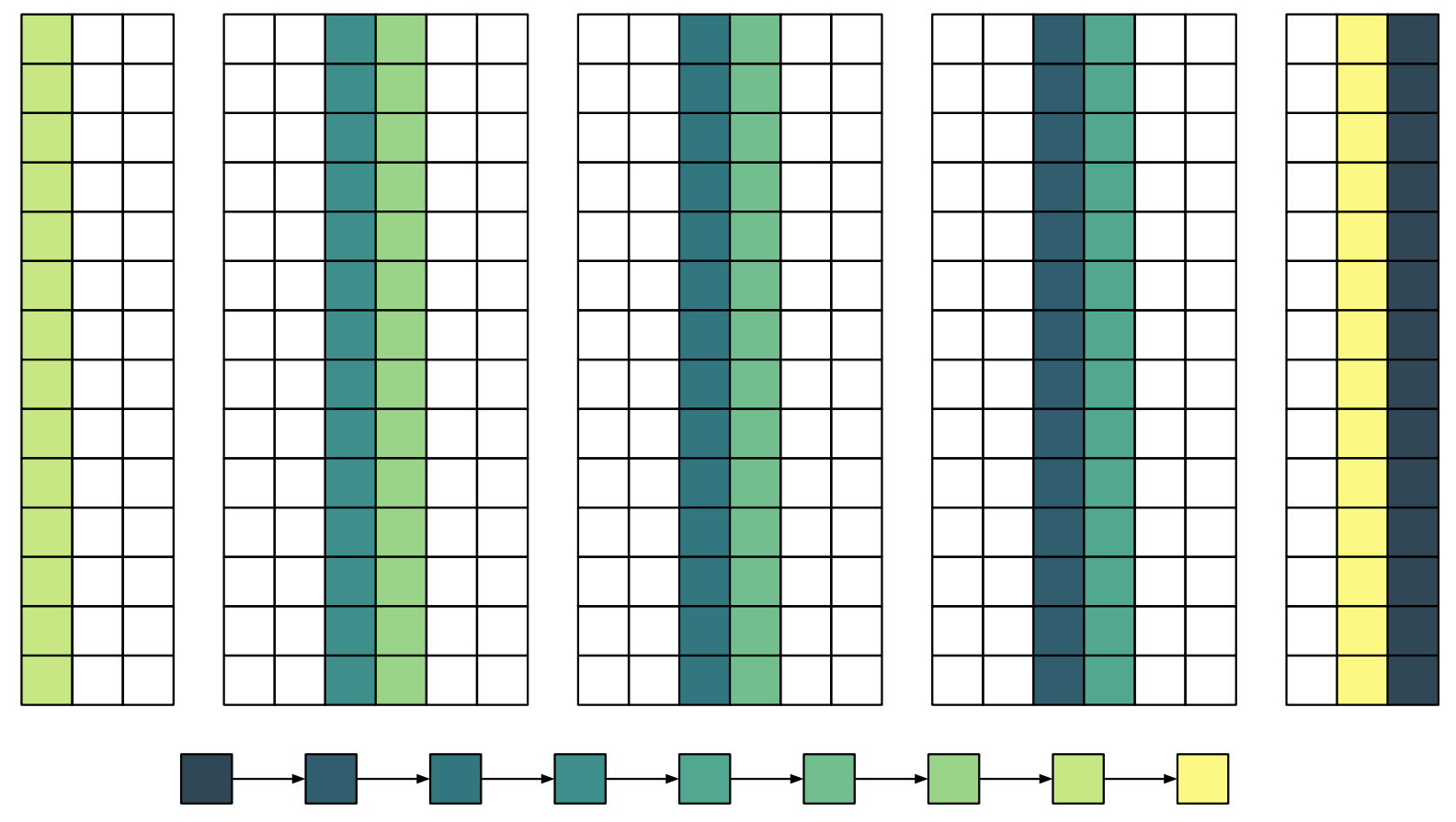}}
\hfil
    \subfloat[Parallel Sequential Steffen \label{PS-Steffen}]{\includegraphics[width=0.47\linewidth]{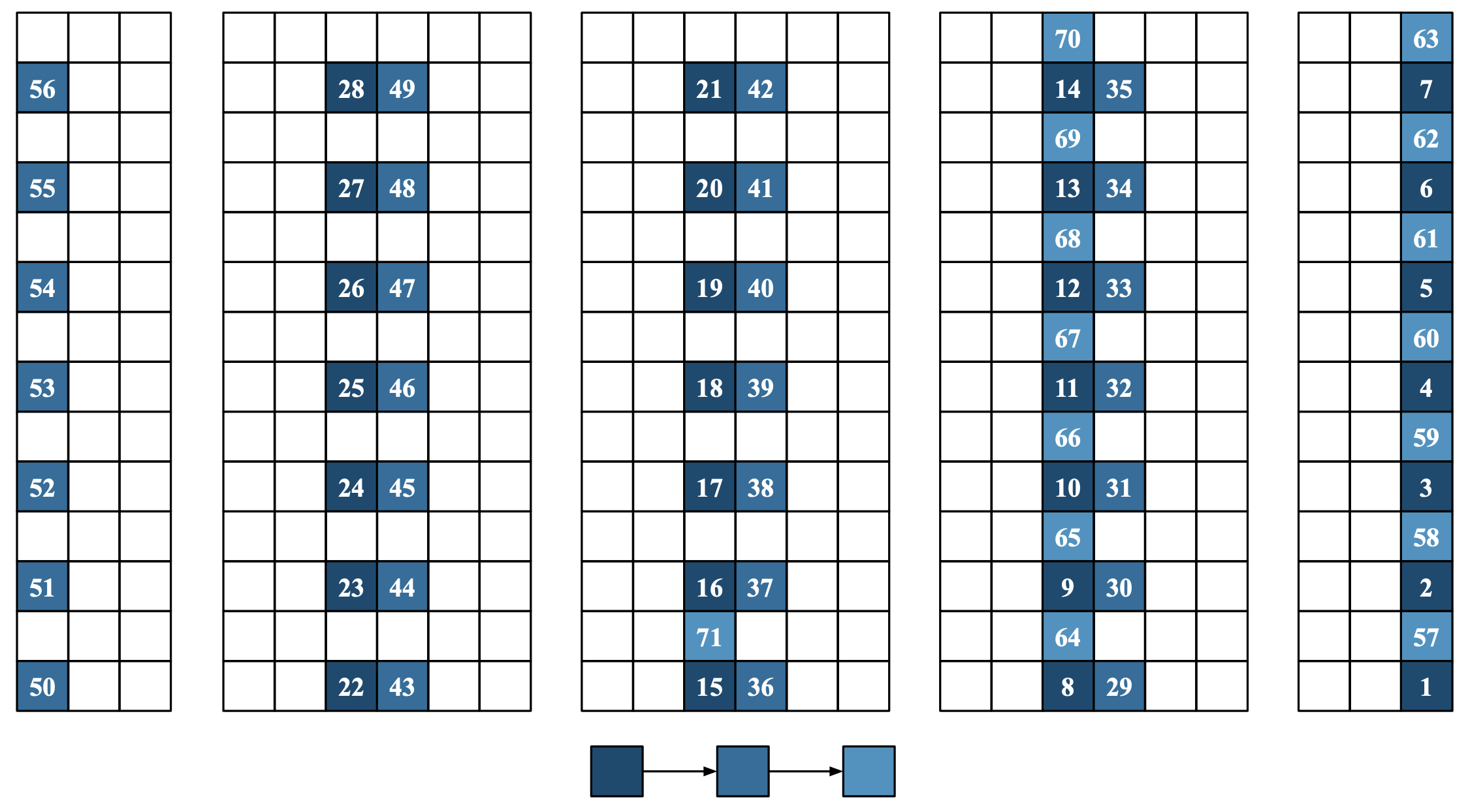}}

    \caption{Visualization of different boarding methods}
    \label{fig:boarding-methods}
    
\end{figure}

\end{enumerate}

All methods above are \textit{strict} in their parallelization in the sense that consecutive passengers always enter separate aisles. We will look into the effect of random parallelization where passengers from a given group could have a seat in any aisle (a more realistic implementation of the method). Those results will be presented in section \ref{subsec:random-vs-deterministic-parallelization}.

\section{Methodology}\label{sec:methodology}

We build an agent-based model in which every passenger represents one agent with individual properties. All passengers follow the same algorithm for determining their action at every time step, and for every time step the model iterates over each agent, computes their respective actions, and then updates them. The program stops when all passengers have entered their seat. To account for the random variables, we execute ensemble runs and average across the realizations.\\

We use Repast Simphony 2.9.1, an open-source platform for agent-based modeling to design and execute the model \citep{repastsimphony}.  For model simulations with some random variables, we measure the averages using ensembles runs of between $10^2$ and $10^3$ realizations, depending on the observed variance of the realizations. This was perceived to give the best trade-off between precision and execution time, a trade-off previously observed by \citep{Schultz2017b}. The time step in the model was taken to be 1 second.\\

Passengers walk directly to the aisle nearest their seat, meaning that no passenger would ever have to move past more than $6/2 = 3$ seats to arrive at their assigned location. In reality, it would be possible for passengers to optimize the boarding procedure somewhat by entering the row of their seat from another aisle, if the aisle closest to their seat was very crowded. The passengers cannot walk past each other and require at least one empty space in front of them to be able to walk forward---except the corners, where the passengers require a corner to be completely clear before they enter it.
\\

For each method, the passengers are ordered and then enter the airplane in that order, with an adjustable time interval between each passenger.  For boarding methods with some randomness in their order, the line of passengers is shuffled by randomly generating the order in which they enter. Each passenger is assigned a walking speed that can be adjusted to any value between 0.0 and 1.0 row per time step.  Every passenger can have carry-on luggage, and if so they are given a positive integer value deciding how much time it will take them to stow their luggage. When stowing their luggage, passengers require at least one of the spaces in the aisle directly adjacent to them to not be occupied by another passenger, to give them enough space to maneuver - this too is a variable which can be changed. Every passenger stows their luggage directly above the row of their respective seat, as soon as they get enough space around them to do so.\\

The model also takes into account the effect of the window, middle, and aisle seating, with an extra time penalty added if a passenger has to seat themselves at the window seat in a row that has already been filled at the middle and/or the aisle seat. The time values for the duration of this seat shuffling were taken from the excellent field trial conducted by \citet{Schultz2018a}.\\

The model can also be tweaked to put different physical constraints on the passengers. As mentioned above, the number of extra spaces required for them to stow their luggage can be adjusted from 0 to 4 adjacent rows (this space requirement is assumed to be constant for all passengers).  Second, the passengers walking speed can be decreased by a constant coefficient between 0.0 and 1.0 in the corners of the airplane.  Third, the extra time it takes for passengers to start after they have stopped and have space to walk again can be set to any real value greater than or equal to 0.

\section{Human Variable Analysis}\label{sec:human-variable-analysis}
By varying each adjustable parameter one at a time, while keeping all other parameters constant, we examined how the boarding time for each method depends on these parameters. In each of these simulations, the parameters of the controlled variables were set to the values defined in table \ref{table_parameters}, apart from two key differences. Firstly, we chose a higher passenger entrance rate of 1 passenger every 2 seconds. This creates higher congestion in the aircraft during boarding, and hence provides a more critical scenario for the different boarding methods to deal with, making it easier to distinguish between them. Figure \ref{Time gap} highlights this, as further discussed below. Secondly, we chose a constant luggage stowage time of 15 seconds, to reduce uncertainty in the model. \\

\begin{figure}[htp]
\centering
    \subfloat[Corner walking speed coefficient \label{Cornering}]{\includegraphics[width=1.0\linewidth]{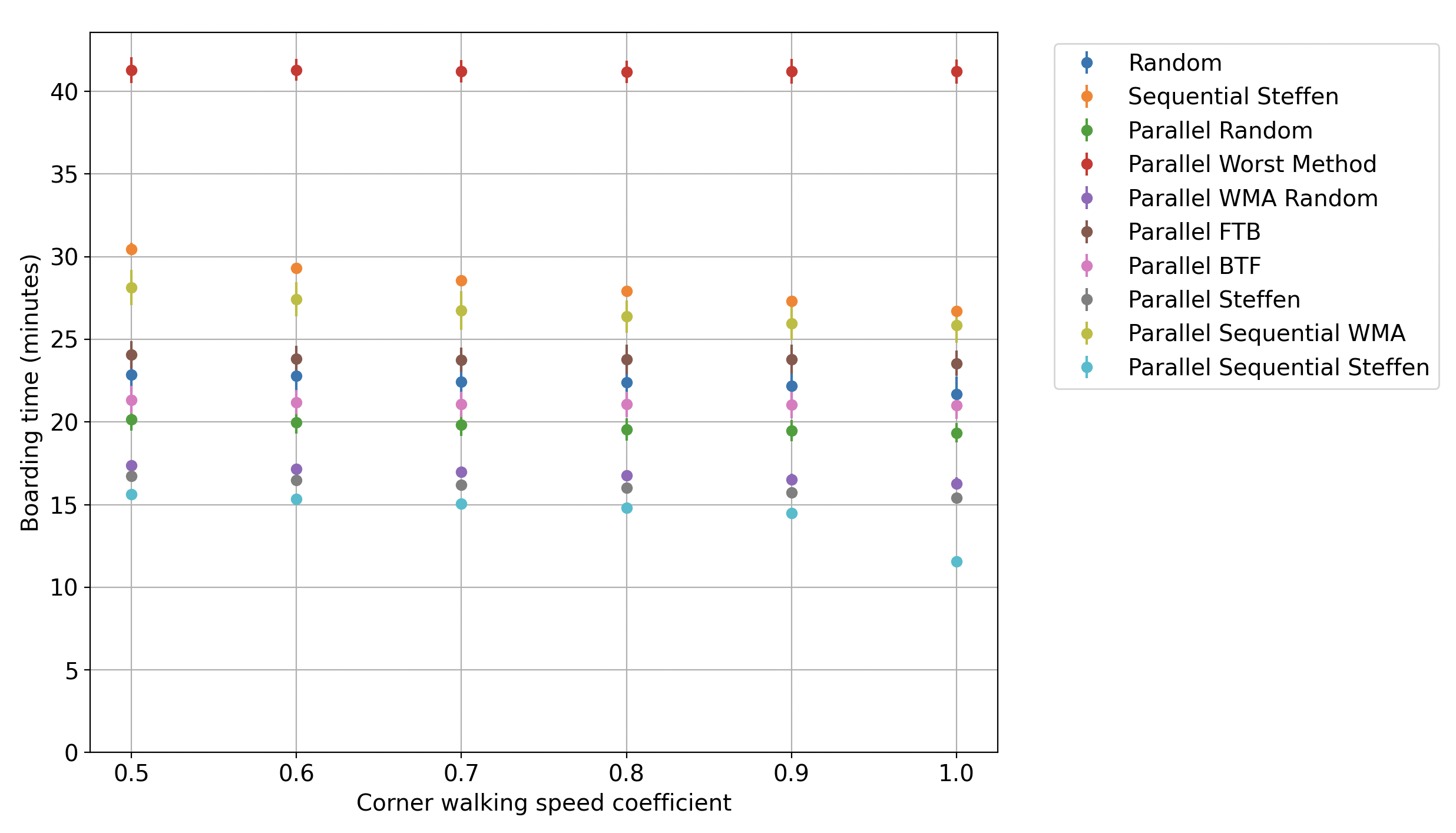}}
\vfil
    \subfloat[Walking restart delay \label{starting time}]{\includegraphics[width=1.0\linewidth]{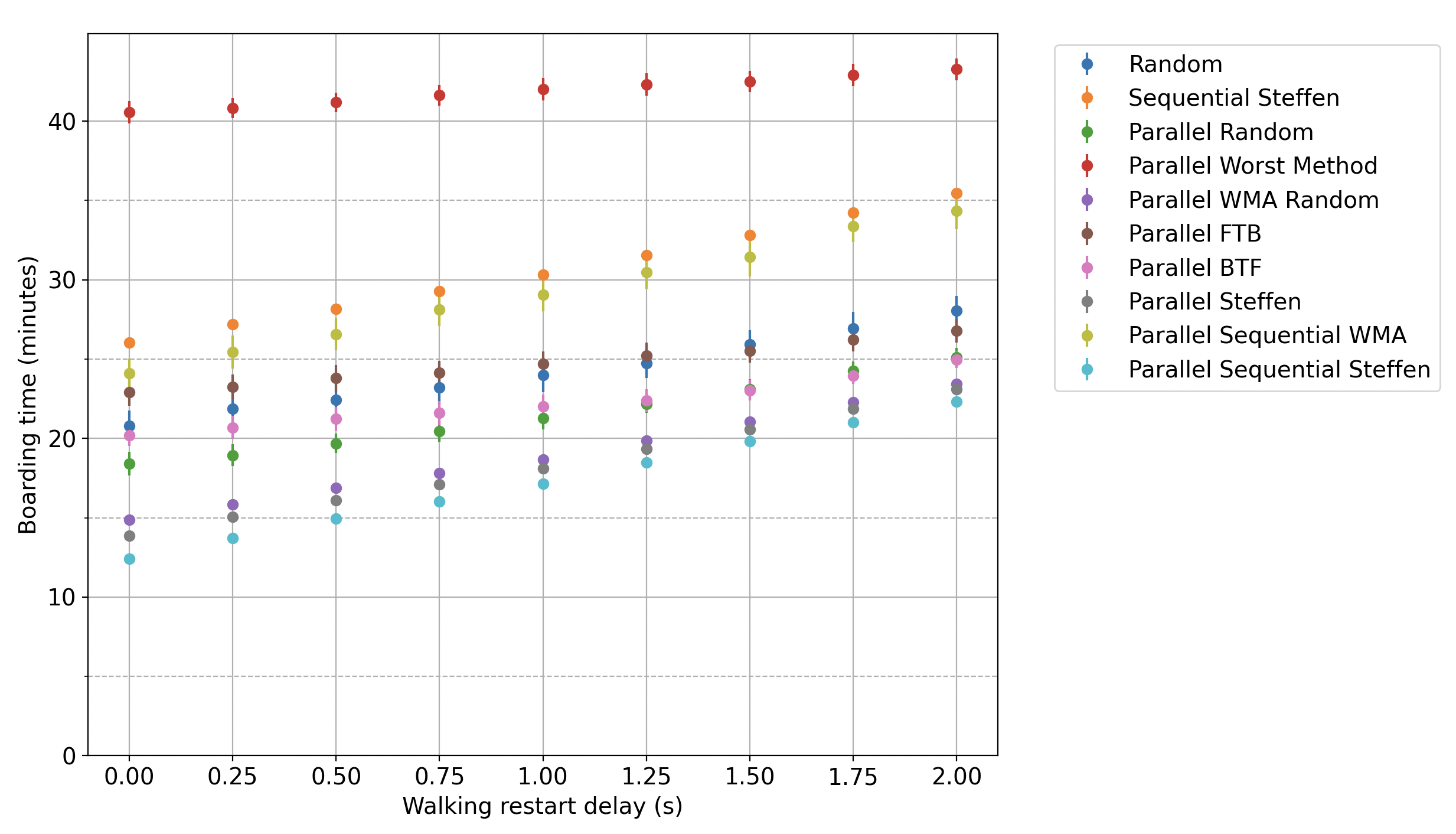}}
\end{figure}
 
\begin{figure}[htp]
\ContinuedFloat

    \subfloat[Time interval between consecutive passengers entering (inverse passenger entrance rate) \label{Time gap}]{\includegraphics[width=1.0\linewidth]{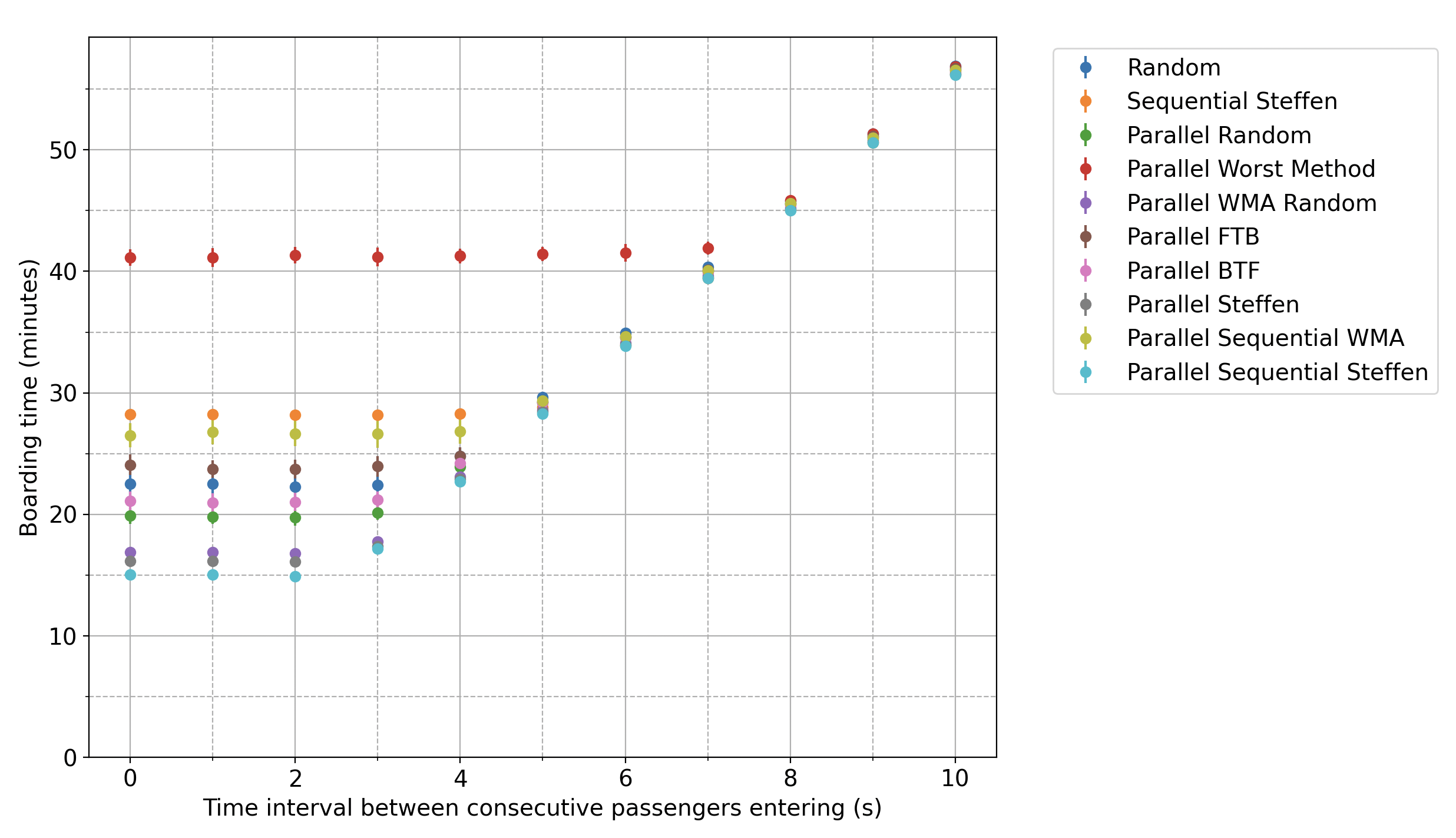}}
    \label{fig:human-variables-time-gap}
\centering
    \subfloat[Luggage stowage time \label{Luggage time}]{\includegraphics[width=\linewidth]{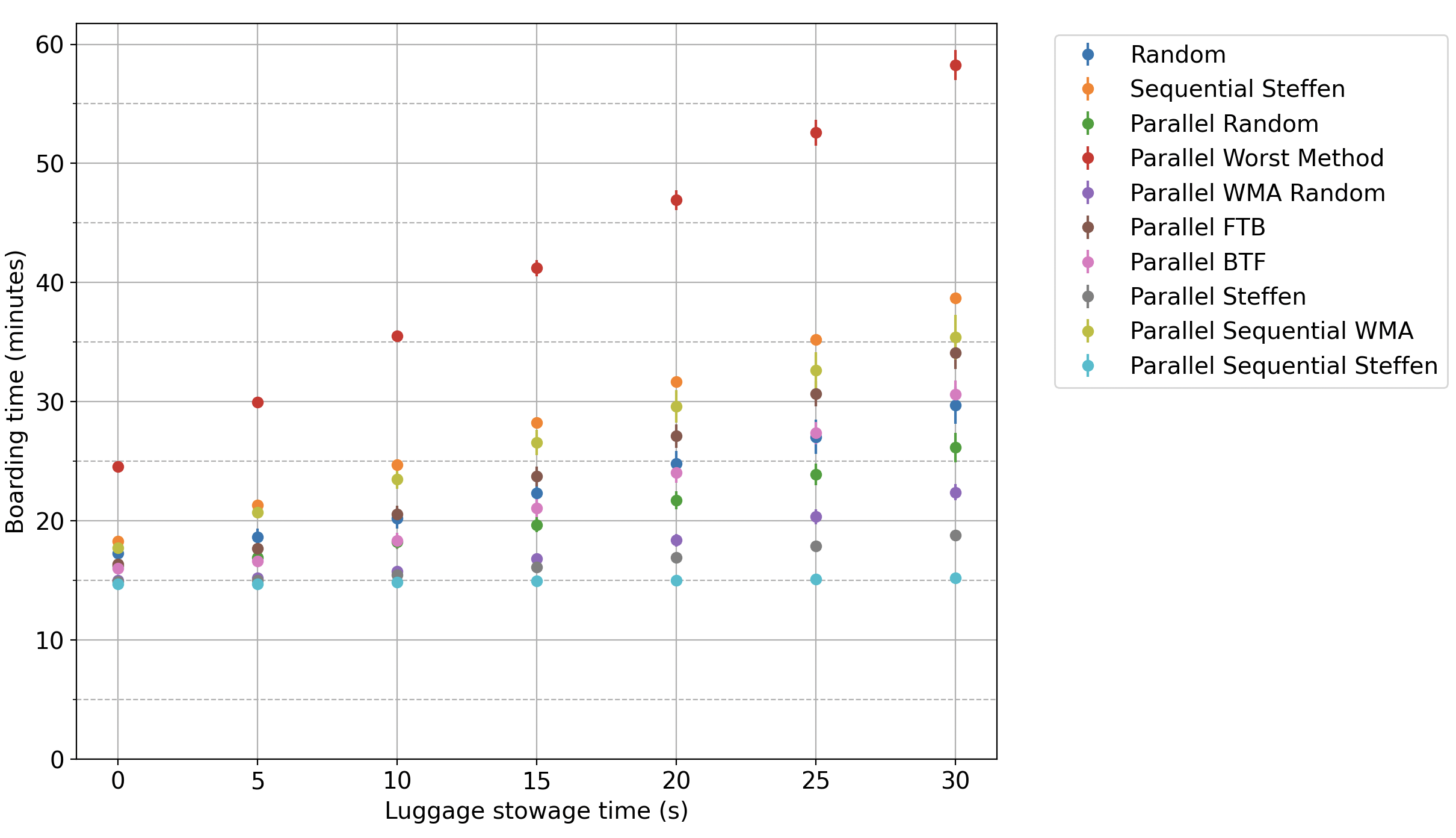}}
\vfil
\end{figure}
\begin{figure}[htp]
\ContinuedFloat

    \subfloat[Standard deviation of luggage stowage time \label{Stowage time standard eviation}]{\includegraphics[width=\linewidth]{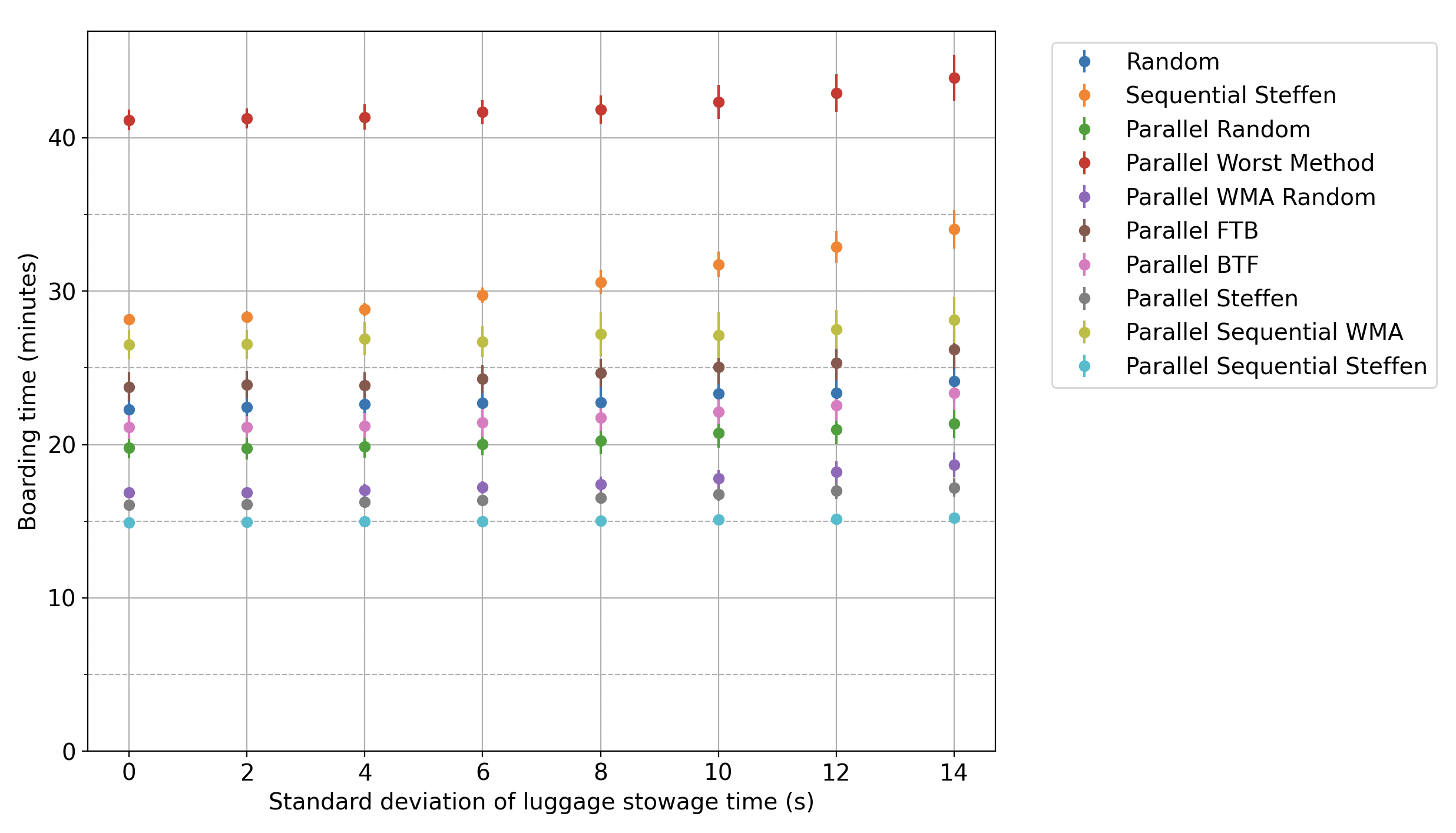}}

\caption{Analysis of the response of the different boarding methods to variations in the human (passenger) variables in a 4-aisle aircraft.}\label{fig:human_variables}
\end{figure}

Figure \ref{Cornering} shows the effect of changing the relative decrease in speed of passengers in the corners, compared to their speed in the aisles. Unsurprisingly, most methods suffer only a marginal increase in boarding time as the walking speed around corners is decreased. Notably, the Parallel Sequential Steffen method - representing the most optimized boarding method given the constraints - becomes slower by a factor of 1.25 as cornering speed is reduced from 1.0 to 0.9. Analyzing the model runs individually, we found that this effect occurs because of the added 0.5 second delay for passengers to restart walking after stopping, which becomes particularly detrimental to the rigid Steffen method, in this case. As corner speed is reduced, this leads to some passengers occasionally having to stop in the corners to wait for the passenger in front of them to exit the corner. As the passengers which stopped then starts walking again, an additional 0.5 second delay is introduced to the collective boarding time. For the other methods, this is a negligible delay, but for the optimized Parallel Sequential Steffen method this delay breaks the otherwise seamless flow of passengers into the aisle, and contributes with a significant increase in boarding time. This point further highlights the idealized constraints necessary for the Steffen method to be drastically faster than other methods.\\

As mentioned in section \ref{sec:methodology}, the walking restart delay is the time it takes a passenger to start moving again after they have stopped and a space has cleared for them to start moving again. Initially, the effect that adjusting the walking restart delay has on the boarding time was thought to be proportional to the number of aisle interferences, and so the methods should react differently depending on how many interferences they create. However, as seen in figure \ref{starting time} it seems that even the methods that have near zero aisle interferences, e.g. the Steffen family of methods, still experienced a significant increase in boarding time as the walking restart delay was increased, and overall the variation in the relative increase was quite small. We believe this to have occurred as aisle interferences were induced by the increased walking restart delay, causing more of a blockage in the aisle as passengers waited longer to start walking again. The hypothesized effect of induced aisle interferences could be of similar proportion for all methods, regardless of their initial number of aisle interferences, hence the similar response to the change in the parameter values. The effect of even a relatively small starting delay of 1 second is quite significant across all boarding methods, leading to an average increase in boarding time of about 18\%, and  starting delay of 2 seconds leads to a boarding time increase of 40\%.\\

Previous publications show that as the rate at which passengers are fed into the airplane decreases (meaning that the time gap between them increases), the different boarding methods eventually converge, as seen in figure \ref{Time gap}.  This occurs because eventually the passengers enter with so much time between them that each passenger has time to stow their luggage and sit before the next one arrives---meaning that the number of aisle collisions will approach zero regardless of method. At a rate of 1 passenger every 8 seconds, all methods are nearly identical in speed. Notably, when the passenger entrance rate is higher than the rate at which a given method has  approached the asymptotic limit, all boarding methods stay almost completely constant in speed even as the passenger rate is lowered. The implication is that as long as the passenger entrance rate is sufficiently high for a given method, there is no incentive for airlines to further increase the entrance rate, as this will yield no decrease in boarding time.\\

The effect of modulating the luggage stowing time is seen in Figure \ref{Luggage time}. In this case, the boarding times for all methods increase linearly with the starting and luggage time, however with a marked variation in the slope. Linearity is to be expected, as each method has a constant total number of passengers stowing their luggage, and so increasing the stowage time for all of them should tend to produce a linear increase in the boarding time. The difference in the slope arises because the different methods produce varying degrees of aisle interferences. For instance, the Parallel Sequential Steffen method shows a negligible increase (increasing only by a factor of 1.03) in boarding time with an increase in the stowage time because the aisle interferences remain at zero regardless of the stowage time.  By contrast, the Parallel Worst Method shows a marked increase by a factor of 2.4 in boarding time as the stowage time is increased because there is a maximum number of aisle interferences. In the limit as luggage stowage time compared to walking time increases, the Steffen family of methods relative advantage grows.\\

To see how the methods would react to increasing variance in the luggage time of the passengers, each passengers luggage time was selected from a normal distribution with mean 15 and with different standard deviations.  When applicable, negative values were set to zero.  This shifted the mean of the distribution by a value on the order of $10^{-1}$, which we deemed insignificant.  Increasing the standard deviation had a marginal effect on all boarding methods (see figure \ref{Stowage time standard eviation}). The Parallel Sequential Steffen method, which typically appears as the least robust to randomness, was the least affected of all methods. This is because the method eliminates all aisle interferences caused by other passengers stowing their luggage. Hence, if one passenger takes longer time (within reason) to stow their luggage, it does not hinder any other passengers from getting to their seat, causing no disruption to the boarding process.

\section{Analysis of Multiple Aisles}\label{sec:analysis-of-multiple-aisles}
With the description of the Flying Wing aircraft in section \ref{sec:methodology}, there are three variables relating to the structure of the plane's seating plan that could be changed: 1) the number of aisles, 2) the number of rows in each aisle, and 3) the number of seats per row.  The Flying Wing model, with its multiple aisles, adds a new complication to the boarding process.  We include here an examination of the effect of the number of aisles on the overall boarding process. The same model parameters were used as in section \ref{sec:human-variable-analysis}.\\

\begin{figure}[H]
\centering
    \subfloat[1 aisle \label{flyingwing1aisle}]{\includegraphics[width=0.25\linewidth, angle = 270]{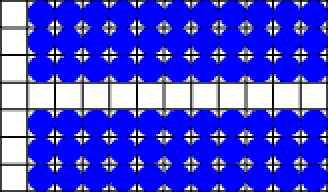}}
\vfil
    \subfloat[10 aisles \label{flyingwing10aisles}]{\includegraphics[width=0.9\linewidth]{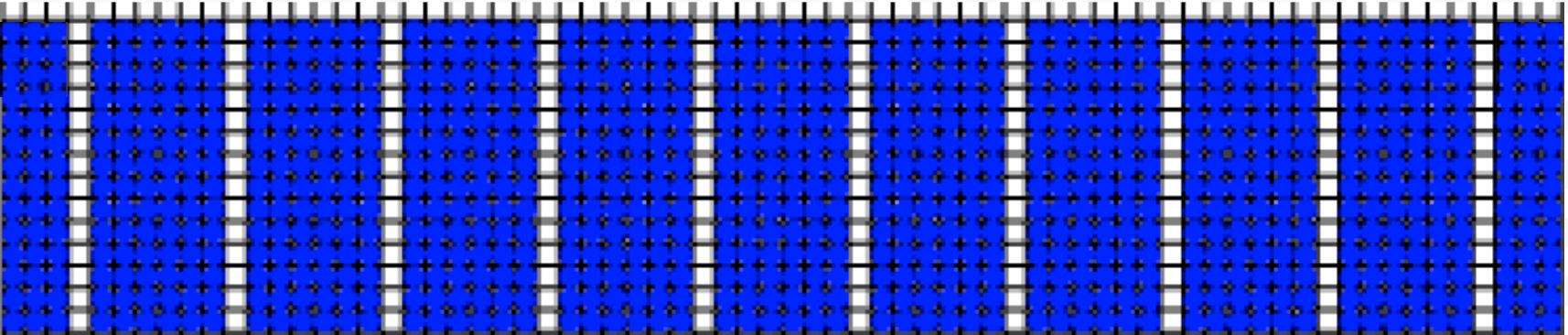}}

    \caption{Aircraft with a) 1 aisle, and with b) 10 aisles, with a constant number of rows.}
    \label{fig:flying-wing-different-aisles}
\end{figure}

\begin{figure}[H]
    \centering
    \includegraphics[scale=0.52]{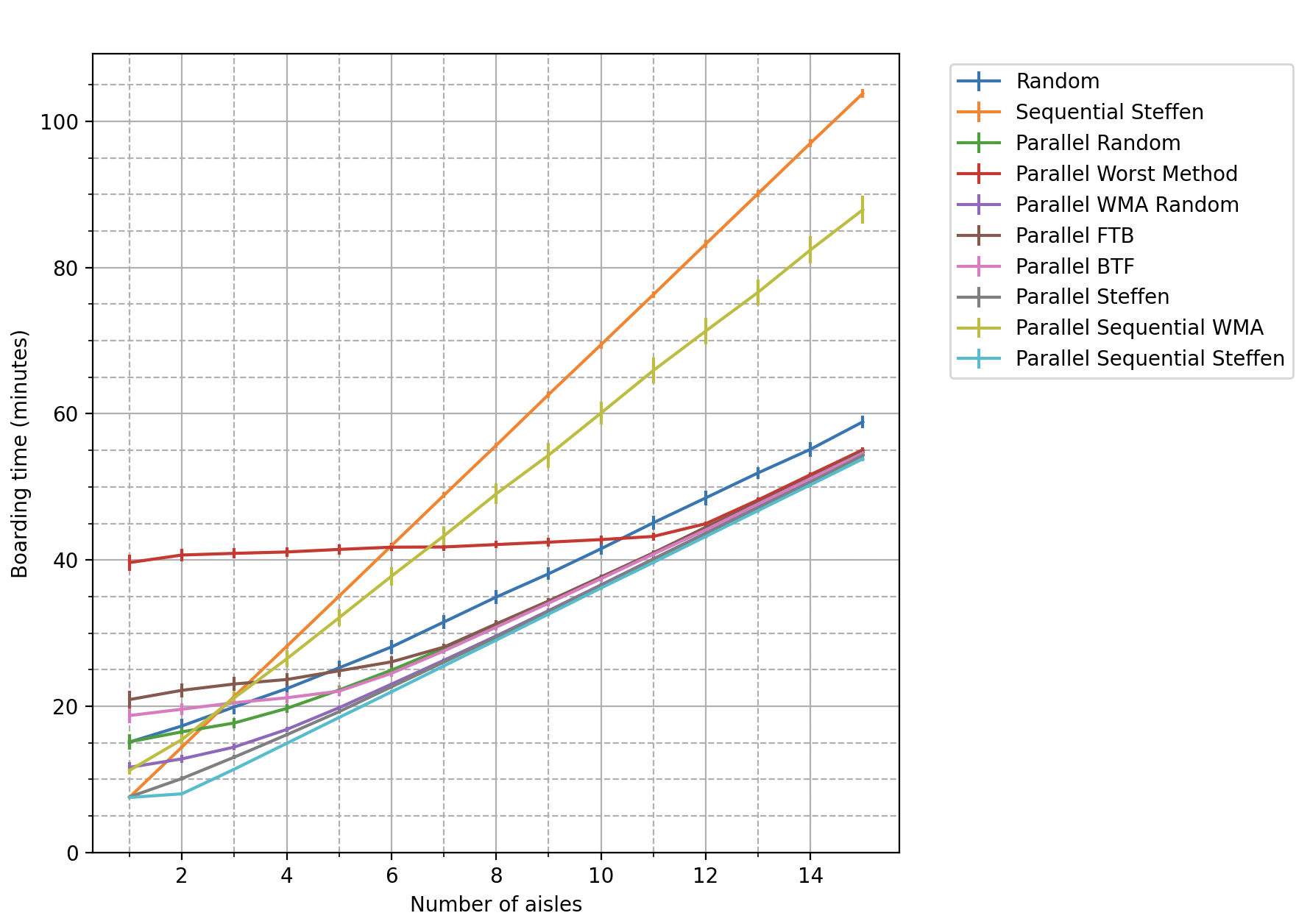}
    \caption{Boarding time (in minutes) vs number of aisles for various boarding methods}
    \label{fig:boarding-time-vs-aisles}
\end{figure}

We investigate how the speed, and the relative advantage or disadvantage of different boarding methods are affected by the introduction of multiple aisles.  To do this, we varied the number of aisles from 1 to 15, as shown in figure \ref{fig:flying-wing-different-aisles}.  For each value, we simulated boarding with each method, and the results are shown in figure \ref{fig:boarding-time-vs-aisles}. For this simulation, the number of rows in each aisle were set to 14 (with exception of the first and last segment with 11 rows in each) and the number of seats per row at three (meaning blocks of six rows per seat in the middle of the airplane). The variables for the passengers were kept constant at their assigned standard values (see section 5.2).\\

Of special interest is the difference between the sequential and parallel methods.  Recalling from section \ref{subsec:boarding-methods}, parallel methods are methods in which successive passengers are sent into different aisles.  By contrast, sequential methods send successive passengers into the same aisle, and then move onto the next aisle after the first one has been completely filled.  As an aircraft increases the number of aisles, parallel methods become increasingly faster compared to sequential methods, as they provide an extra way of parallelizing the boarding procedure.

Furthermore, these methods will, for a large enough number of aisles converge asymptotically to the optimum method, as there will be so much time between the nth and (n+1)th passenger entering a given aisle that the first passenger in a given aisle will have time to be seated before the next passenger in that aisle arrives.  The speed of this convergence varies between different boarding methods, as seen in figure \ref{fig:boarding-time-vs-aisles}.\\

We found that the Parallel Sequential Steffen Method converges to the fastest boarding time immediately, as it is specifically constructed to eliminate all aisle interferences.  The method maximizing aisle interferences, Parallel Worst Method should be the slowest to converge out of all parallel methods, and is here displayed as a theoretical maximum value for conversion time, at roughly 10 aisles.  That is, in our model, the time it takes for a passenger to stow their luggage and seat themselves is roughly the time it takes for them to walk past nine aisles in the Flying Wing layout.\\

Some of the methods, e.g. the Sequential Steffen method, presented in the figure are not parallel, and so do not converge---meaning that they do not approach the optimal boarding time, given by the Parallel Sequential Steffen method---as it doesn't utilize the many aisles to board them simultaneously, instead boarding them one by one. This causes the boarding time to increase linearly with the number of aisle segments.  Similarly, the Parallel Sequential WMA Random method also does not converge, as even though it does rely on some parallel processes in boarding, it still puts the nth and (n+1)th passenger in the same aisle after each other (although their seat row is randomly assigned), meaning that the sought after effect of the nth passenger having seated themselves before the (n+1)th passenger arrives never occurs.\\

Inspecting the other methods, we note that at one aisle, equivalent to normal one-aisle aircraft (with only 14 rows, however), the Parallel or Parallel Sequential Steffen method is roughly 2 to 3 times faster than other relevant methods, but already at four aisles (a reasonable number for a Flying Wing aircraft) it is only up to 1.6 times faster. Thus, employing multi-aisle aircraft could significantly improve boarding speed, without forcing airlines to resort to strict ordering methods.\\

We note that Figure \ref{fig:boarding-time-vs-aisles} looks similar to \ref{Time gap}.  This is because they both show roughly the same effect.  As the number of aisles increase, the time interval between passengers sitting next to each other also increases (assuming a parallel boarding method), so the specifics of the method become gradually less important.\\

\section{Results}\label{sec:results}
\subsection{Comparing a single aisle with a 4-aisle Flying Wing aircraft}\label{subsec:simulation-results}

The different boarding models were applied to a Flying Wing design with four aisles, as well as to a traditional one-aisle aircraft for comparison.  The number of passengers in each airplane was kept constant by changing the number of rows, and so the three aircraft had 56 and 14 rows in each aisle respectively. In total, each simulated aircraft had 336 passengers. To compare these different aircraft, we chose the following model parameters, most of which are presented in the table below. For luggage stowage time, walking speed, seat shuffling time, and passenger entrance rate there are empirically observed values from the literature (see table). However, for corner walking speed factor, walking restart delay, stowage space and carry-on luggage percentage we chose values based on what we believed to be representative assumptions of the average passenger.


\begin{center}
    \begin{tabular}{
    >{\centering\arraybackslash}m{15em}
    >{\centering\arraybackslash}m{10em}}
    \hline \textbf{Variable} &  \textbf{Value}\\
    \hline Luggage Time & Empirical distribution as found in \citep{Schultz2018a}, approximately equal to a Weibull distribution with parameters $\beta = 16$ seconds, $\alpha = 1.7 $ \citep{Schultz2018a} \\
    \hline Passenger entrance rate & $1/3.7$ passengers/seconds \citep{Schultz2018a}\\
    \hline Walking Speed & 1 row/s \citep{Schultz2018a}\\
    \hline Carry-on luggage per passenger & 0.75 \citep{Hutter2019} \\ 
    \hline Stowage Space & 1 Row \\
    \hline Walking restart delay & 0.5 seconds \\
    \hline Cornering Speed Factor & 0.75 \\
    
    \hline
    \end{tabular}\label{table_parameters}
\end{center}

\begin{figure}[H]
\centering
    \subfloat[1 aisle \label{1 aisle}]{\includegraphics[width=0.65\linewidth]{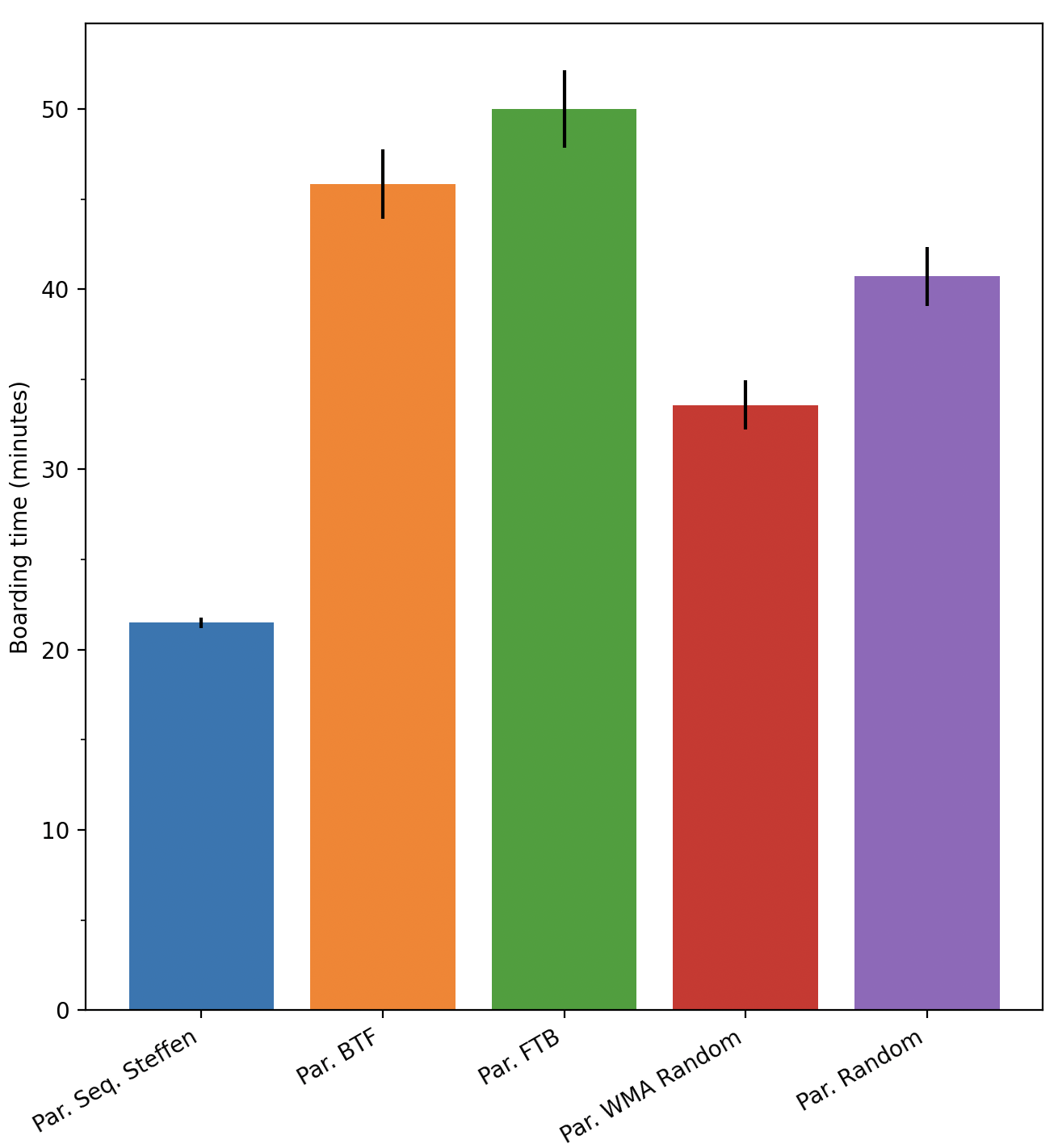}}
\end{figure}
\begin{figure}[H]
\ContinuedFloat
\centering
    \subfloat[4 aisles \label{4 aisles}]{\includegraphics[width=0.65\linewidth]{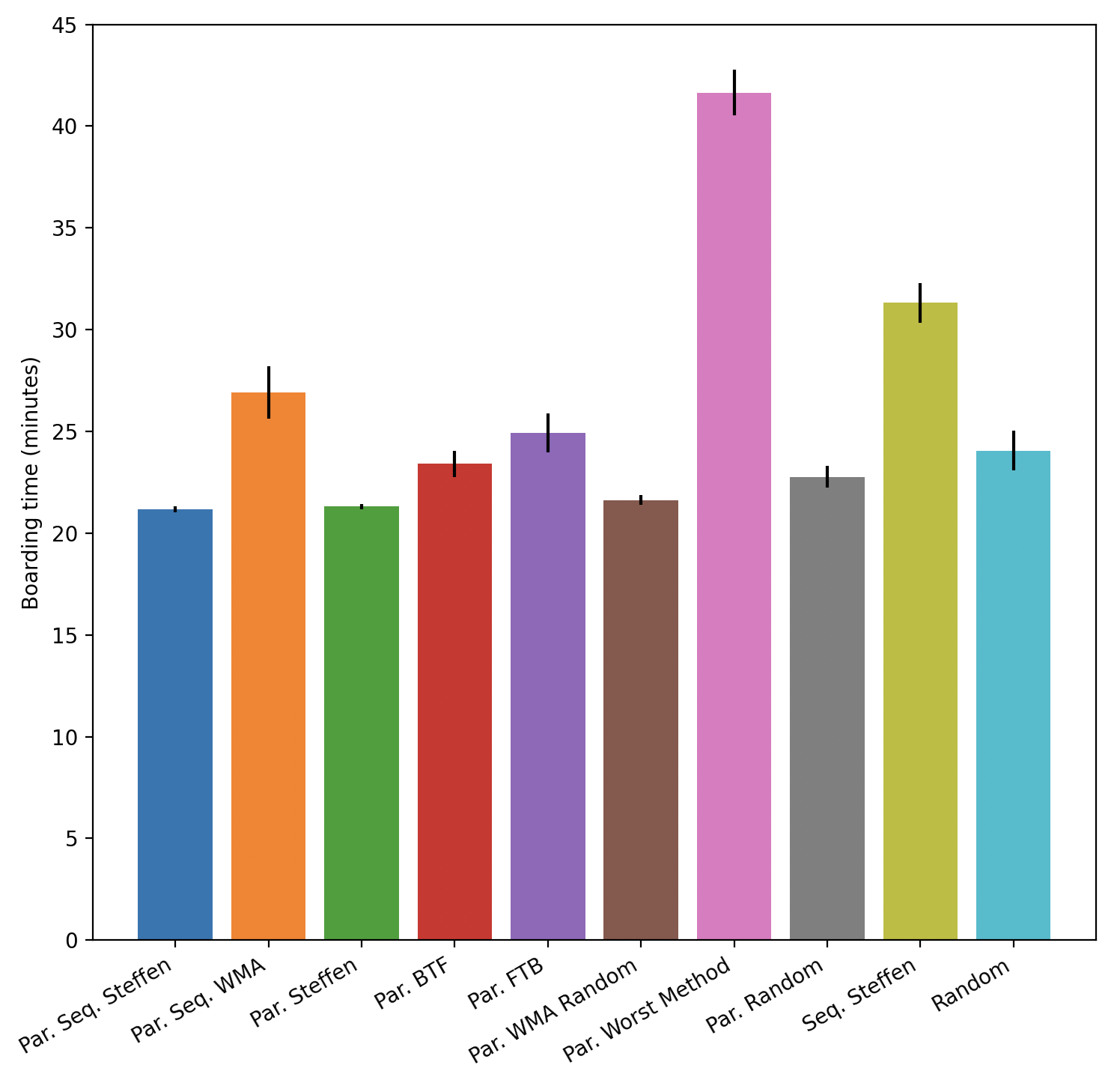}}
    
    \caption{Boarding times of different boarding methods in a single-aisle, double-aisle, and 4-aisle (Flying Wing) aircraft model}
    \label{fig:final-results}
\end{figure}

Figure \ref{fig:final-results} shows how the effects of parallelization can work to decrease the relative disadvantage of traditional methods compared to the Steffen method.  As seen in \ref{4 aisles}, the best method (Parallel Sequential Steffen) is approximately equal in speed to Parallel WMA Random in the Flying Wing aircraft with four aisles, compared to the one aisle aircraft where it is 1.6 faster.  Further, Parallel BTF is only around 1.1 times slower than the Parallel Sequential Steffen, while being 2.1 slower in the one-aisle aircraft. \\

It may be seen that the worst methods, like the Parallel Worst Method, show the largest relative decrease in boarding time from the narrow aircraft to the Flying Wing. Thus, the range of boarding times for the same number of passengers is significantly smaller for the Flying Wing aircraft than the one-aisle and two-aisle aircraft.  The boarding times decrease by a factor of approximately 0.5 BTF and FTB, and around 0.7 for WMA, if passengers are allowed to board in parallel in four aisles instead of all in one.\\

Similar to results from previous papers \citep {Belgium2002} \citep{Steffen2008}, of the parallel methods the Steffen method is consistently the fastest, followed by WMA, Random, Back-to-front, and lastly Front-to-back.  Interestingly, since the Random method does not systematically utilize parallel boarding, like the parallel methods, it becomes a relatively less enticing choice of method as the number of aisles increase.  For instance, as can be seen in figure \ref{fig:boarding-time-vs-aisles}, with four aisles, Random and Parallel BTF are nearly the same value, but as the number of aisles increases, the latter overtakes the former as the faster method. The Sequential Steffen Method (predictably) increases linearly with the number of aisles and is thus not a good choice for a multi-aisle aircraft, as shown in section \ref{sec:analysis-of-multiple-aisles}.

\subsection{Random vs deterministic parallelization}\label{subsec:random-vs-deterministic-parallelization}
In the results above, all parallel methods employed have been deterministically parallel, meaning that for an aircraft with $m$ aisles, for each set of $m$ passengers entering the airplane after each other, precisely one passenger will enter into each aisle. While this is shown to lead to a significant decrease in boarding times, it is an impractical sorting method which would likely lead to large increases in the presorting time at the gate, thus negating the benefits.\\

To increase practicality for airlines, they might opt instead to parallelize their boarding procedure randomly across the aisles. For instance, for Parallel BTF, this would mean that all passengers seated in the back rows board first (regardless of the aisle of their respective seats). This will indeed slow down boarding procedure compared to fully deterministic parallelization, but it may simplify the staging of the passengers and therefore be worth the additional time. Here we quantify the size of this difference in boarding time. Once again, to distinguish more clearly between different boarding methods, the passenger entrance rate was increased to one passenger every 2 seconds in these simulations, and the luggage stowage time set to a constant 15 seconds.

\begin{figure}[H]
    \centering
    \includegraphics[scale=0.4]{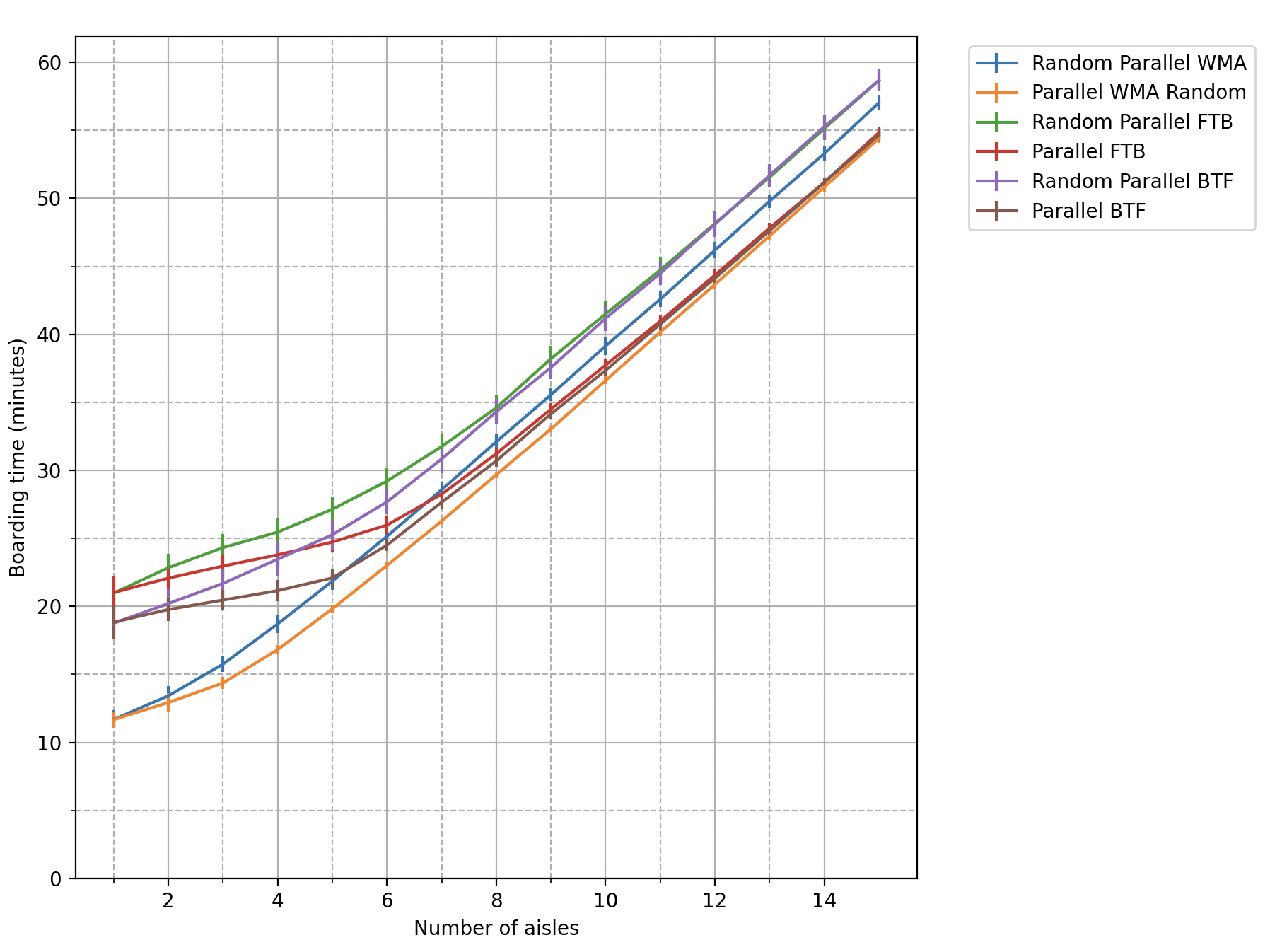}
    \caption{Comparison of boarding time (in seconds) vs number of aisles for traditional boarding methods with both random and deterministic parallelization}
    \label{fig:boarding-time-vs-aisles-random-vs-deterministic}
\end{figure}

\begin{figure}[H]
    \centering
    \includegraphics[scale=0.45]{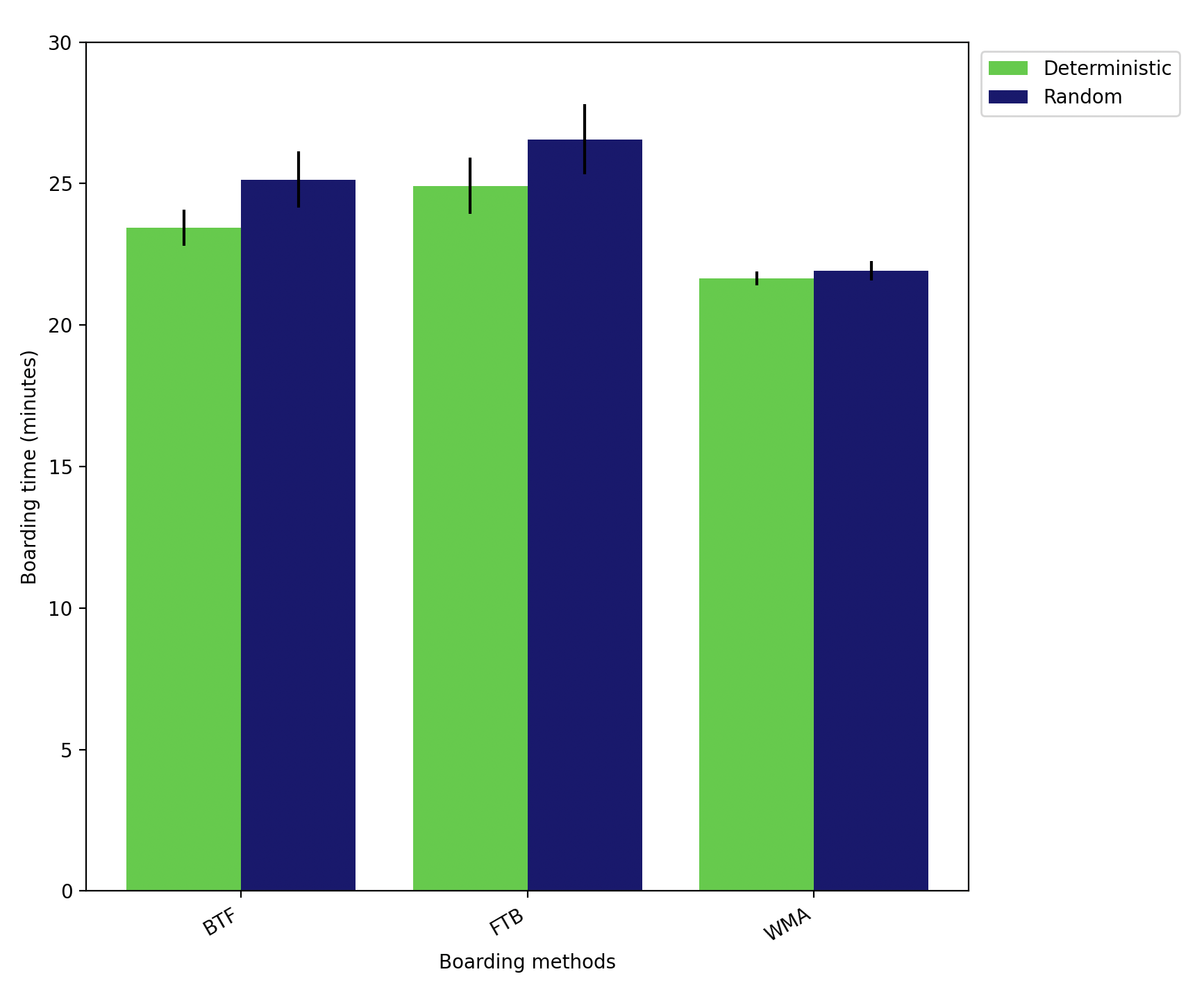}
    \caption{Comparison of boarding time at four aisles for traditional boarding methods with both random and deterministic parallelization}
    \label{fig:boarding-time-vs-aisles-4-aisles-random-vs-deterministic}
\end{figure}

Figure \ref{fig:boarding-time-vs-aisles-random-vs-deterministic} shows that as the number of aisles increase, the randomly parallel methods do converge to a different optimal boarding time than the deterministically parallel methods.  The effect that causes the convergence of the methods is less pronounced for the methods that are not completely parallelized, since there will still be instances where adjacent passengers will board the plane into the same aisle.  Thus, the difference between the random and deterministic version of each method seems to grow as the number of aisles increases, approaching a constant value after a sufficient number of aisles. \\

Comparing the two versions for the specific case of four aisles, the fiducial case investigated in this paper, we see slight differences between the methods (see Figure \ref{fig:boarding-time-vs-aisles-4-aisles-random-vs-deterministic}). In this simulation run, we return to the empirically observed passenger entrance rate and luggage stowage time, to provide the most operationally useful results. The difference in boarding time is roughly 7\% for BTF, 6\% for FTB and just 1\% for WMA.  For our purposes, the use of deterministically parallel methods in this paper can be used as a good indication of the relative strength of different boarding methods, although randomly parallel methods may be slightly slower than their deterministic alternatives.

\section{Discussion and Implications for Management}\label{sec:disc-management}

Apart from the results presented in this paper, the investigative approach used here will be useful for optimizing the aircraft interior design and boarding strategies.  We showed how the boarding process is affected by the aisle and row layout and how that layout performs with parallel versus sequential boarding groups.  We also explored the impact of changes in human variables on the boarding process.  As shown in figure \ref{Luggage time}, the average luggage time of passengers has a significant impact on the total boarding time for Back-to-front and Front-to-back boarding. Taking measures to ensure the speed of luggage stowing should therefore be considered, for instance by having part of the airline crew assist passengers in stowing their luggage.\\

The analysis of the impacts of multiple aisles shows that a multi-aisle aircraft such as the Flying Wing would present an excellent opportunity for airline managers to decrease the boarding time for their passengers without resorting to boarding methods that are overly challenging to implement. Simply by boarding the plane by groups based on rows, and thus filling up the aisles in parallel, boarding times can decrease significantly. Furthermore, this improvement would not come at an increase in presorting time at the gate, as it would be no more difficult to implement than standard Back-to-front boarding. As parallel boarding methods are implemented, while maintaining a fair degree of simplicity, airlines may improve the passenger experience by making the boarding process faster without imposing additional complications for passengers. \\

The methods outlined in this paper provide a variety of options to explore when considering how to effectively board an aircraft with more than one aisle. The fastest and slowest parallel methods (the Parallel Steffen and the Parallel Worst Method) provide a theoretical maximum and minimum boarding time for a given seating architecture. These results can be used for evaluating a given method by comparing it with the extreme cases.

\section{Conclusion}
\label{sec:conclusion}
In this work, we found that the introduction of multiple aisles into aircraft seating design, as proposed for the Flying Wing, offers the possibility of significantly decreasing the boarding time per passenger if methods that board the aisles in parallel are employed.  The different variations of the Steffen method, adapted for the Flying Wing, are still faster than other alternatives, yet its relative advantage compared to the other parallel methods decreases as the number of aisles increases.\\

The Steffen method's relative advantage decreases, in part, because there are more aisles---leading to a lower passenger density in each aisle and fewer aisle interferences.  However, because we used a fixed number of passengers, by increasing the number of aisles, the number of rows per aisle also decreases.  This also has an effect on the relative superiority of the Steffen Method, which gains in superiority as the number of rows increases.\\

We note that the parallel methods investigated in this paper have been optimized to feed passengers into the aisles one at a time, starting from the aisle most distant from the door.  In reality, this might be difficult to ensure, and so we proposed a more feasible, random parallel method. We found that this practical implementation of parallel boarding still provides very fast boarding, with only a minor decrease in speed compared to the idealized, deterministic parallel boarding.\\

We conclude that the Flying Wing aircraft presents a good opportunity for airlines to significantly speed up their boarding process while still retaining traditional boarding methods, adapted to the multi-aisle design. This will provide economic opportunity for the airlines, while simultaneously improving passenger experience. Although the boarding process is not the main reason for developing the this airplane design, it is a benefit for airlines who eventually choose to acquire Flying Wing designs as part of their fleet.\\





\bibliography{cas-refs.bib}

\begin{thebibliography}{}

\bibitem[KLM, 2020]{KLM}
 (2020).
\newblock Klm and tu delft present successful first flight flying-v.
\newblock Accessed on Wednesday, April 24, 2024.

\bibitem[Audenaert et~al., 2009]{Audenaert2009}
Audenaert, J., Verbeeck, K., and Berghe, G. (2009).
\newblock Multi-agent based simulation for boarding.
\newblock In {\em Proceedings of the 21st Benelux Conference on Artificial Intelligence, Eindhoven, The Netherlands}, pages 29--30. Citeseer.

\bibitem[Bachmat, 2019]{Bachmat2019}
Bachmat, E. (2019).
\newblock Airplane boarding meets express line queues.
\newblock {\em European Journal of Operational Research}, 275(3):1165--1177.

\bibitem[Bachmat et~al., 2009]{Bachmat2009}
Bachmat, E., Berend, D., Sapir, L., Skiena, S., and Stolyarov, N. (2009).
\newblock Analysis of airplane boarding times.
\newblock {\em Operations Research}, 57:499--513.

\bibitem[Bazargan, 2011]{Bazargan2011}
Bazargan, M. (2011).
\newblock A linear programming approach for wide-body two- aisle aircraft boarding strategy.
\newblock {\em International journal of operations and quantitative management}, 17(3):193--210.

\bibitem[Benad, 2015]{Benad2015}
Benad, J. (2015).
\newblock The flying v - a new aircraft configuration for commercial passenger transport.

\bibitem[Cook and Tanner, 2011]{Cook2011EuropeanAD}
Cook, A. and Tanner, G. (2011).
\newblock European airline delay cost reference values.

\bibitem[Cotfas et~al., 2020]{Milne2020Covid}
Cotfas, L.-A., Delcea, C., Milne, R., and Salari, M. (2020).
\newblock Evaluating classical airplane boarding methods considering covid-19 flying restrictions.
\newblock {\em Symmetry}, 12:1087.

\bibitem[Erland et~al., 2022]{Erland2022}
Erland, S., Bachmat, E., and Steiner, A. (2022).
\newblock Let the fast passengers wait: Boarding an airplane takes shorter time when passengers with the most bin luggage enter first.
\newblock {\em European Journal of Operational Research}.

\bibitem[Erland et~al., 2019]{Erland2019}
Erland, S., Kaupu\ifmmode~\check{z}\else \v{z}\fi{}s, J., Frette, V., Pugatch, R., and Bachmat, E. (2019).
\newblock Lorentzian-geometry-based analysis of airplane boarding policies highlights ``slow passengers first'' as better.
\newblock {\em Phys. Rev. E}, 100:062313.

\bibitem[Erland et~al., 2021]{Erland2021}
Erland, S., Kaupu\ifmmode~\check{z}\else \v{z}\fi{}s, J., Steiner, A., and Bachmat, E. (2021).
\newblock Lorentzian geometry and variability reduction in airplane boarding: Slow passengers first outperforms random boarding.
\newblock {\em Phys. Rev. E}, 103:062310.

\bibitem[Ferrari and Nagel, 2005]{Ferrari2005}
Ferrari, P. and Nagel, K. (2005).
\newblock Robustness of efficient passenger boarding strategies for airplanes.
\newblock {\em Transportation Research Record}, 1915(1):44--54.

\bibitem[Fuchte, 2014]{Fuchte2014}
Fuchte, J. (2014).
\newblock Enhancement of aircraft cabin design guidelines with special consideration of aircraft turnaround and short range operations.
\newblock {\em DLR Deutsches Zentrum fur Luft- und Raumfahrt e.V. - Forschungsberichte}, pages 1--141.

\bibitem[Giitsidis and Sirakoulis, 2016]{Giitsidis2016}
Giitsidis, T. and Sirakoulis, G. (2016).
\newblock Modeling passengers boarding in aircraft using cellular automata.
\newblock {\em IEEE/CAA Journal of Automatica Sinica}, 3:365--384.

\bibitem[Hutter et~al., 2019]{Hutter2019}
Hutter, L., Jaehn, F., and Neumann, S. (2019).
\newblock Influencing factors on airplane boarding times.
\newblock {\em Omega}, 87:177--190.

\bibitem[Jaehn and Neumann, 2015]{AB}
Jaehn, F. and Neumann, S. (2015).
\newblock Airplane boarding.
\newblock {\em European Journal of Operational Research}, 244(2):339--359.

\bibitem[Kierzkowski and Kisiel, 2017]{KIERZKOWSKI2017}
Kierzkowski, A. and Kisiel, T. (2017).
\newblock The human factor in the passenger boarding process at the airport.
\newblock {\em Procedia Engineering}, 187:348--355.
\newblock TRANSBALTICA 2017: TRANSPORTATION SCIENCE AND TECHNOLOGY: Proceedings of the 10th International Scientific Conference, May 4–5, 2017, Vilnius Gediminas Technical University, Vilnius, Lithuania.

\bibitem[Liang and Bazargan, 2016]{Bazargan2016}
Liang, Y. and Bazargan, M. (2016).
\newblock A simulation study on boarding and deplaning utilizing two-doors for a narrow body aircraft.
\newblock {\em International Journal of Aviation Systems, Operations and Training}, 3:25--35.

\bibitem[Marelli et~al., 1998]{Marelli1998}
Marelli, S., Mattocks, G., and Merry, R. (1998).
\newblock The role of computer simulation in reducing airplane turn time.
\newblock {\em Boeing Aero Magazine}, 1.

\bibitem[Milne and Kelly, 2014]{NewMethod}
Milne, R.~J. and Kelly, A.~R. (2014).
\newblock A new method for boarding passengers onto an airplane.
\newblock {\em Journal of Air Transport Management}, 34:93--100.

\bibitem[Milne and Salari, 2016]{MILNE2016}
Milne, R.~J. and Salari, M. (2016).
\newblock Optimization of assigning passengers to seats on airplanes based on their carry-on luggage.
\newblock {\em Journal of Air Transport Management}, 54:104--110.

\bibitem[Neumann, 2019]{NEUMANN2019}
Neumann, S. (2019).
\newblock Is the boarding process on the critical path of the airplane turn-around?
\newblock {\em European Journal of Operational Research}, 277(1):128--137.

\bibitem[North et~al., 2013]{repastsimphony}
North, M., Collier, N., Ozik, J., Tatara, E., Altaweel, M., Macal, C., Bragen, M., and Sydelko, P. (2013).
\newblock Complex adaptive systems modeling with repast simphony.
\newblock {\em Complex Adaptive Systems Modeling}, 1.

\bibitem[Notomista et~al., 2016]{NOTOMISTA2016}
Notomista, G., Selvaggio, M., Sbrizzi, F., {Di Maio}, G., Grazioso, S., and Botsch, M. (2016).
\newblock A fast airplane boarding strategy using online seat assignment based on passenger classification.
\newblock {\em Journal of Air Transport Management}, 53:140--149.

\bibitem[Nugroho and Asrol, 2022]{Nugroho2022}
Nugroho, A.~A. and Asrol, M. (2022).
\newblock The impact of effectiveness of luggage arrangement on the airplane passengers' boarding process.
\newblock {\em Periodica Polytechnica Transportation Engineering}, pages 369--386.

\bibitem[Nyquist and McFadden, 2008]{GeneralStudy}
Nyquist, D.~C. and McFadden, K.~L. (2008).
\newblock A study of the airline boarding problem.
\newblock {\em Journal of Air Transport Management}, 14(4):197--204.

\bibitem[Schultz, 2017a]{Schultz2017b}
Schultz, M. (2017a).
\newblock Dynamic change of aircraft seat condition for fast boarding.
\newblock {\em Transportation Research Part C: Emerging Technologies}, 85:131--147.

\bibitem[Schultz, 2017b]{Schultz2017}
Schultz, M. (2017b).
\newblock Faster aircraft boarding enabled by infrastructural changes.
\newblock In {\em 2017 Winter Simulation Conference (WSC)}, pages 2530--2541.

\bibitem[Schultz, 2018a]{Schultz2018_58}
Schultz, M. (2018a).
\newblock Fast aircraft turnaround enabled by reliable passenger boarding.
\newblock {\em Aerospace}, 5(1).

\bibitem[Schultz, 2018b]{Schultz2018a}
Schultz, M. (2018b).
\newblock Field trial measurements to validate a stochastic aircraft boarding model.
\newblock {\em Aerospace}, 5(1).

\bibitem[Schultz et~al., 2013]{Schultz2013}
Schultz, M., Kunze, T., and Fricke, H. (2013).
\newblock Boarding on the critical path of the turnaround.

\bibitem[Steffen, 2008]{Steffen2008}
Steffen, J.~H. (2008).
\newblock Optimal boarding method for airline passengers.
\newblock {\em Journal of Air Transport Management}, 14(3):146–150.

\bibitem[Steffen and Hotchkiss, 2012]{STEFFENExperiment}
Steffen, J.~H. and Hotchkiss, J. (2012).
\newblock Experimental test of airplane boarding methods.
\newblock {\em Journal of Air Transport Management}, 18(1):64--67.

\bibitem[Tang et~al., 2012]{TANG2012}
Tang, T.-Q., Wu, Y.-H., Huang, H.-J., and Caccetta, L. (2012).
\newblock An aircraft boarding model accounting for passengers’ individual properties.
\newblock {\em Transportation Research Part C: Emerging Technologies}, 22:1--16.

\bibitem[van~den Briel et~al., 2005]{vandenBriel2005}
van~den Briel, M. H.~L., Villalobos, J.~R., Hogg, G.~L., Lindemann, T., and Mulé, A.~V. (2005).
\newblock America west airlines develops efficient boarding strategies.
\newblock {\em Interfaces}, 35(3):191--201.

\bibitem[{Van Landeghem} and Beuselinck, 2002]{Belgium2002}
{Van Landeghem}, H. and Beuselinck, A. (2002).
\newblock Reducing passenger boarding time in airplanes: A simulation based approach.
\newblock {\em European Journal of Operational Research}, 142(2):294--308.

\bibitem[Wamelink, 2021]{Wamelink2021}
Wamelink, L. (2021).
\newblock Flying-v interior: Floorplan design for improved passenger comfort.
\newblock Master's thesis, TU Delft.

\bibitem[Willamowski and Tillmann, 2022]{Willamowski2022}
Willamowski, F. J.~L. and Tillmann, A.~M. (2022).
\newblock Minimizing airplane boarding time.
\newblock {\em Transportation Science}, 56(5):1196–1218.

\bibitem[Xin-hui et~al., 2017]{IEEEBoarding}
Xin-hui, R., Xi-yu, Z., and Yang, J. (2017).
\newblock The passenger boarding strategy of two-aisle aircraft under different situations.
\newblock In {\em 2017 9th International Conference on Modelling, Identification and Control (ICMIC)}, pages 110--115.

\end{thebibliography}






\end{document}